\documentclass[twocolumn,showpacs,amsmath,amssymb,aps,prb,superscriptaddress]{revtex4-1}

\usepackage[]{graphicx}
\usepackage[]{verbatim}
\usepackage[]{color}
\usepackage{overpic}
\usepackage{rotating}
\usepackage{mathtools}
\usepackage{appendix}
\usepackage{ulem}
\usepackage[margin=0.6in]{geometry}

\usepackage[]{hyperref}

 \hypersetup{
 colorlinks = true,
 linkcolor = blue,
 urlcolor = magenta,
 citecolor = magenta}

\usepackage{array}
\newcolumntype{L}[1]{>{\raggedright\let\newline\\\arraybackslash\hspace{0pt}}m{#1}}
\newcolumntype{C}[1]{>{\centering\let\newline\\\arraybackslash\hspace{0pt}}m{#1}}
\newcolumntype{R}[1]{>{\raggedleft\let\newline\\\arraybackslash\hspace{0pt}}m{#1}}
\newcolumntype{N}{@{}m{0pt}@{}}

\makeatletter
\newsavebox{\@brx}

\newcommand{\llangle}[1][]{\savebox{\@brx}{\(\m@th{#1\langle}\)}%
  \mathopen{\copy\@brx\mkern2mu\kern-0.8\wd\@brx\usebox{\@brx}}}
\newcommand{\rrangle}[1][]{\savebox{\@brx}{\(\m@th{#1\rangle}\)}%
  \mathclose{\copy\@brx\mkern2mu\kern-0.8\wd\@brx\usebox{\@brx}}}

  \newcommand{\lllangle}[1][]{\savebox{\@brx}{\(\m@th{#1\langle}\)}%
  \mathopen{\copy\@brx\copy\@brx\mkern4mu\kern-0.7\wd\@brx\usebox{\@brx}}}
\newcommand{\rrrangle}[1][]{\savebox{\@brx}{\(\m@th{#1\rangle}\)}%
  \mathclose{\copy\@brx\copy\@brx\mkern4mu\kern-0.7\wd\@brx\usebox{\@brx}}}

\graphicspath{{./}{./}}

\begin{document}
\title{Microscopic mechanism for higher-spin Kitaev model}
\author{P. Peter Stavropoulos}
\affiliation{Department of Physics and Center for Quantum Materials, University of Toronto, 60 St.~George St., Toronto, Ontario, M5S 1A7, Canada}
\author{D. Pereira}
\affiliation{Department of Physics and Center for Quantum Materials, University of Toronto, 60 St.~George St., Toronto, Ontario, M5S 1A7, Canada}
\author{Hae-Young Kee}
\affiliation{Department of Physics and Center for Quantum Materials, University of Toronto, 60 St.~George St., Toronto, Ontario, M5S 1A7, Canada}
\affiliation{Canadian Institute for Advanced Research, Toronto, Ontario, M5G 1Z8, Canada}
\email{hykee@physics.utoronto.ca}

\begin{abstract}
The spin  S=$\frac{1}{2}$ Kitaev honeycomb model has attracted significant attention, since 
emerging candidate materials have provided a playground to test 
non-Abelian anyons.
The Kitaev model with higher spins has also been theoretically studied, as it may offer another path to a quantum spin liquid.
However, a microscopic route to achieve higher spin Kitaev models in solid state materials has not been rigorously derived.
Here we present a theory of the spin S=1 Kitaev interaction in two-dimensional edge-shared octahedral systems.
Essential ingredients are strong spin-orbit coupling in anions and strong Hund's coupling in transition metal cations.
The S=1 Kitaev and ferromagnetic Heisenberg interactions are generated from superexchange paths.
Taking into account the antiferromagnetic Heisenberg term from direct-exchange paths, the Kitaev interaction dominates the physics of S=1 system.
Using exact diagonalization technique, we show a finite regime of S=1 spin liquid in the presence of the Heisenberg interaction.
Candidate materials are proposed, and generalization to higher spins is discussed.
\end{abstract}
\maketitle

\textit{Introduction} -- 
Highly entangled quantum spin liquids provide exotic phenomena including fractional excitations.\cite{Balents2010NA,Zhou2017RMP}
Among several proposed quantum spin liquid models, 
an exactly solvable model is a bond-dependant interaction of spin S=$\frac{1}{2}$ on the two-dimensional (2D) honeycomb lattice proposed by Kitaev.\cite{Kitaev2006AoP}
The ground state of the S=$\frac{1}{2}$ Kitaev model offers non-Abelian anyons under a magnetic field. Recently the smoking-gun evidence of 
such particles was supported by the half-integer quantized thermal Hall conductivity in $\alpha$-RuCl$_3$\cite{Matsuda2018NA},
making it the most promising candidate to display Kitaev physics.

\begin{figure}[!ht]

  \centering
 \begin{overpic}[width=0.97\linewidth]{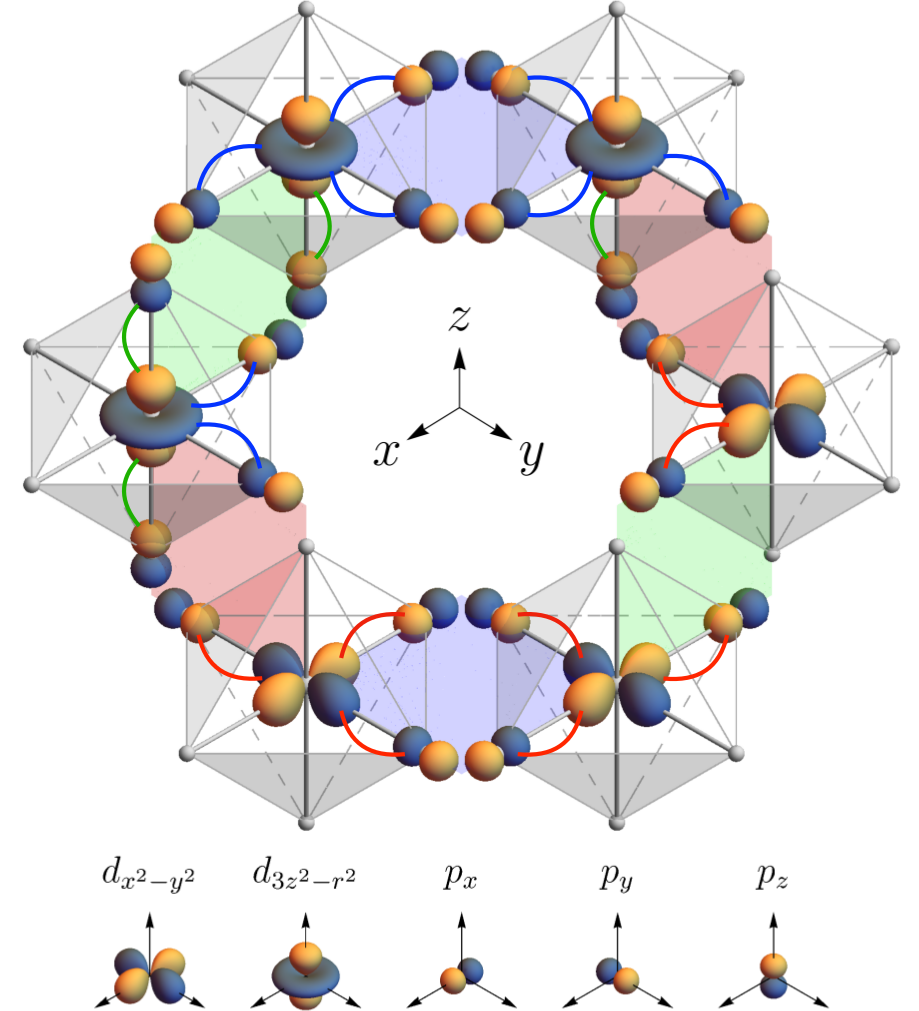}
\end{overpic}
  \caption{Indirect hopping integrals 
  between M and A sites
  are denoted by the colored curve lines. The red, green, and blue colors represent $t_1$, $t_2$ and $t_3$ respectively, and the sign of the hopping integrals is ignored for simplicity. 
  The M sites with e$_g$ orbitals are located in the center of each octahedral cage formed by A sites occupied by three p orbitals. 
  Kitaev bond dependent interactions  X-, Y-, and Z-bond are respectively represented by red, green, and blue shaded regions.
  For clarity, every A site is drawn by two separated A sites to represent different hopping contributions from
  different p and e$_g$ orbtials. The global coordinates of 
  $x$-, $y$-, and $z$-axes are shown in the center of the honeycomb plane.}
    \label{fig:hoppings}
\end{figure}

Along with the rapid progress on the S=$\frac{1}{2}$ Kitaev spin liquids in solid state materials \cite{Khaliullin2009PRL,
Chaloupka2010PRL,
Gegenwart2012PRL,
Taylor2012PRL,
ChaloupkaPRL2013,
Balents2014ARoCMP,
Rau2014PRL,
YJKim2014PRB,
Rau2014ARXIV,
Sears2015PRB,
HSKim2015PRB,
Sandilands2016PRB,
Sandilands2015PRL,
Banerjee2016NAM,
Rau2016ARoCMP,
Johnson2015PRB,
HSKim2016PRB,
Cao2016PRB,
Janssen2017PRB,
Winter2017IOP}, the theoretical condensed matter physics community has considered a higher spin S
Kitaev model.
A first attempt was made by Baskaran and collaborators.\cite{Baskaran2008PRB} They showed that for arbitrary spin S, localized Z$_2$ flux excitations are present, as 
plaquette operators
can be constructed, and a vanishing spin-spin correlation beyond nearest-neighbors
is found.\cite{Baskaran2007PRL} Unlike the S=$\frac{1}{2}$ model, the higher S Kitaev model is not exactly solvable, and several
numerical studies have been performed.
 In particular, the S=1 Kitaev model has been studied by using exact diagonalization (ED) and  thermal pure quantum (TPQ) techniques
and it was suggested that the ground state of the S=1 Kitaev model may be a gapless spin liquid.\cite{Nasu2018JPSJ} Using high-temperature series expansions
and TPQ, a double peak structure in the specific heat similar to S=$\frac{1}{2}$  and an incipient entropy plateau at value of $\frac{1}{2} {\rm ln} 3$ 
were found in the S=1 model.\cite{Singh2018PRB,Nasu2018JPSJ}.
Dynamics of the classical (S $\rightarrow \infty$) Kitaev spin liquid was also studied and it was suggested that the quantum model can be
understood by fractionalization of magnons in one-dimensional manifolds.\cite{BatistaPRB2017} 
While these theoretical results promote another path to quantum spin liquids, there has been a lack of microscopic routes to achieve spin S Kitaev model in solid state materials. 

In this letter,  we present a way to generate the S=1 bond-dependent Kitaev interaction in 2D Mott insulators with edge-shared octahedra.
Two essential ingredients are strong Hund's coupling among two electrons in e$_g$-orbitals and strong spin-orbit coupling (SOC) at anion sites.
Using a strong coupling expansion, we show
that the bond-dependant interactions are generated via superexchange between two cations with e$_g$ orbitals mediated by anion p orbital electrons with strong SOC.
12- and 18-site ED results of S=1 Kitaev-Heisenberg (KH) model
show a finite regime of the Kitaev spin liquid.  Candidate materials are proposed, and generalization to higher spin bond-dependent
interactions are also discussed. 
                                                
\textit{Microscopic mechanism for S=1 Kitaev model} -- 
We consider a 2D edge-shared octahedral system with two types of atoms. 
The honeycomb (or triangular) network is made of transition metal (M) cations with half filed e$_g$ orbitals such as $d^8$ electronic configuration. 
The anion (A) atoms with fully occupied p orbitals form edge-shared octahedral cages around every M site as shown in Fig. \ref{fig:hoppings}.
The Hamiltonian consists of the on-site interactions $H_0$ and hopping between M and A sites, $H_{\mathrm{kin}}$.
For 3d transition metals, such as Ni$^{2+}$, 
typical energy scales of the hopping parameters are smaller than the energy scales of the on-site $H_0$, which allows the use of standard strong coupling expansion theory.
The on-site Hamiltonian of both M and A sites is described by the Kanamori interaction\cite{Kanamori1963PoTP} and SOC:
\begin{eqnarray}
H_0& =  &  U \sum\limits_{\alpha} n_{\alpha\uparrow} n_{\alpha\downarrow} + \dfrac{U'}{2} \sum\limits_{\substack{\alpha\neq\beta,\\\sigma,\sigma'}} n_{\alpha\sigma} n_{\beta\sigma'} + \lambda \mathbf{l} \cdot \mathbf{s} \\
& & - \dfrac{J_{H}}{2}\sum\limits_{\substack{\alpha\neq\beta,\\\sigma,\sigma'}} c^{\dagger}_{\alpha\sigma}  c^{\dagger}_{\beta\sigma'} c_{\beta\sigma} c_{\alpha\sigma'}
+ J_{H}\sum\limits_{\alpha\neq\beta} c^{\dagger}_{\alpha\uparrow} c^{\dagger}_{\alpha\downarrow} c_{\beta\downarrow} c_{\beta\uparrow}\nonumber,
\label{eq:kanamori}
\end{eqnarray}
where the density operator $n_{\alpha\sigma}$ is given by $ c^{\dagger}_{\alpha \sigma} c_{\alpha \sigma}$, and
$c^{\dagger}_{\alpha\sigma}$ is the creation operator with $\alpha$ orbital and spin $\sigma$. 
$U$ and $U'$ are the intra-orbital and inter-orbital density-density interaction respectively, and $J_{H}$ is the Hund's coupling for the spin-exchange and pair-hopping terms.
Operators $\mathbf{l}$ and $\mathbf{s}$ respectively denote angular momentum and spin for orbital $\alpha$ and spin $\sigma$, and $\lambda$ denotes the strength of SOC.  

In general the competition between the Hund's coupling and SOC leads to a different atomic state.\cite{fazekas1999}
For the M sites with e$_g$ orbitals, the SOC is inactive when the crystal field splitting between t$_{2g}$ and e$_g$ is bigger than the SOC strength.
Here we consider $d^8$ systems, such as Ni$^{2+}$, where t$_{2g}$ orbitals are fully filled, and the crystal field splitting is larger than the SOC. 
In this case, the SOC does not mix the e$_g$ states, as the e$_g$ orbitals are made of the z-component of
angular momentum of $\pm 2$ and 0. In the half-filed e$_g$ orbitals the Hund's coupling selects the total spin S=1 state with energy $U'-J_{H}$. 
On the other hand, for the A sites with p orbitals, the SOC splits the p orbitals into total angular momentum $j=\frac{3}{2}$ and 
$j=\frac{1}{2}$ states. The Hund's coupling for A sites is only relevant for excited states in the perturbation theory.
The full energy spectrum of $H_0$ required for the perturbation theory is listed in Table 1 in the supplementary material (SM).
To differentiate $U$, $U'$, $J_H$ for d and p orbitals, we use subscript $d/p$ for $U_{d/p}$, $U'_{d/p}$,
and $J_{H_{d/p}}$, which refer to the on-site interactions for d/p-orbitals from now on.
Similarly we use $d^\dagger$ and $p^\dagger$ to represent creation operator for d and p orbital respectively.
For SOC, we have only $\lambda_p$ because $\lambda_d$ is inactive in the e$_g$ orbitals when the crystal field splitting is larger than the SOC,
which is the case for 3d systems. 

Let us consider the nearest neighbour (n.n.) hopping parameters between the M and A sites to construct a minimal n.n. spin model. 
Since the p orbitals are fully filled, and e$_g$ orbitals are half-filled, we consider holes rather than electrons.
Then in the ground state of the atomic Hamiltonian $H_0$, there is no hole in the p orbitals while $e_g$ orbitals are half filed.
It is straightforward to build the tight-binding model:
\begin{equation}
H_{\mathrm{kin}} = \sum\limits_{\substack{\langle i,j \rangle\\\sigma}} d^{\dagger}_{i,\alpha\sigma} M^{(i,j)}_{\alpha,\beta} p_{j,\beta\sigma} + h.c.,
\end{equation}
where $d^{\dagger}_{i,\alpha\sigma}$($p^{\dagger}_{j,\beta \sigma}$)
creates one of d(p) orbitals denoted by $\alpha$($\beta$) and spin $\sigma$ on site $i$($j$).
The hopping matrix $M^{(i,j)}$ depends on the $(i,j)$ bond.
As shown in Fig. \ref{fig:hoppings}, there are three distinct hopping integrals $t_1$, $t_2$ and $t_3$ denoted by
the red, blue, and green colored curves respectively.
They appear on different bonds. 
For example, 
\begin{equation}
\begin{array}{l}
\text{Along x-axis}: \ \ t_1 \; d^{\dagger}_{i,x^2-y^2} p_{j,x} - t_2 \; d^{\dagger}_{i,3z^2 - r^2} p_{j,x} + h.c. \\
\text{Along y-axis}:  -t_1 \; d^{\dagger}_{i,x^2-y^2} p_{j,y}  -t_2 \; d^{\dagger}_{i,3z^2-r^2} p_{j,y} + h.c. \\
\text{Along z-axis}: \ \ t_3 \; d^{\dagger}_{i,3z^2-r^2} p_{j,z} + h.c.
\end{array}
\end{equation}
\noindent
All other bond directions are related by symmetry such as mirror symmetry, and the set of tight binding parameters is given in Table 2 in the SM. 
They can be represented by the Slater-Koster parameters\cite{SlaterKoster}, i.e., $t_1=\frac{\sqrt{3}}{2}t_{pd\sigma},\ t_2=\frac{1}{2} t_{pd\sigma},\ t_3=t_{pd\sigma}$
if the perfect cubic symmetry is preserved. 

Treating the tight binding Hamiltonian $H_{\rm kin}$ as a perturbation to the on-site Hamiltonian $H_0$, a 
n.n. spin model for S=1 on the honeycomb lattice with edge-shared octahedra via superexchange processes is determined.
Before we derive the model explicitly, it is straightforward to check that the symmetry of the edge-shared octahedral crystal allows
Heisenberg $J$, Kitaev $K$, and symmetric off-diagonal $\Gamma$ interactions.\cite{Rau2014PRL,Imada2014PRL,VdBrink2014IOP}.

There are several processes that contribute to the spin interaction and we categorize them by the number of holes at a given site.
The one hole processes include intermediate states with one hole at most on any A site and the two hole processes
include intermediate states with two holes on an A site.
In the one hole processes, the SOC $\lambda_p$ generates intermediate states of different energies, depending on whether
the one hole state is $j=\frac{1}{2}$ or $\frac{3}{2}$.
For the two hole process, p orbital Hund's coupling $J_{H_p}$ becomes as important as the SOC, and we will consider
two limits of $J_{H_p} \rightarrow 0$ and $\lambda_p \rightarrow 0$ to show the origin of Kitaev interaction.

Taking into account all possible fourth-order superexchange processes shown in the SM, 
the resulting n.n. spin model consists of the Kitaev and Heisenberg interactions:
\begin{equation}
\label{eq:hamiltonian}
  H_{\langle ij \rangle}^{\gamma} =  K^\gamma S_i^{\gamma} S_j^{\gamma} + J_{\rm ind} {\bf S}_i  \cdot {\bf S}_j,
\end{equation}
where $i,j$ are n.n. sites, and $\gamma$ refers the X-, Y-, and Z-bond type. ${\bf S}$ is the spin 1 operator
and its bond-dependent interaction takes $\gamma={x,y,z}$ spin component. 
The spin components are directed along the cubic axes of the underlying ligand octahedra, so the honeycomb layer lies in a plane perpendicular to the [111] spin direction as shown in Fig. \ref{fig:hoppings}.
Note that $\Gamma$ term is exactly 0 within the fourth-order term, and $J_{\rm ind} = -\frac{1}{2} K^\gamma$.

The expressions of $K^\gamma$ and $J_{\rm ind}$ are presented in the SM and they can be simplified in certain limits.
With the cubic symmetry, i.e., $t^2_1= \frac{3}{4} t_3^2$ and $t^2_2 = \frac{1}{4} t_3^2$, $K^{x/y} = K^z \equiv K$.
When $U_d$ and the atomic potential difference $\Delta$ between the M $(\epsilon_M$) and 
A $(\epsilon_A$) sites (i.e, $\Delta \equiv \epsilon_M - \epsilon_A$) are the largest energy scales, it simplifies to
\begin{equation}
K \sim \frac{3}{2}\lambda_p^2 t_{pd\sigma}^4\left( \frac{1}{\left(2U_d +\Delta\right)^5}+\frac{1}{2U_d\left(2U_d +\Delta\right)^4}\right).
\end{equation}
For a Mott insulator, i.e.,  $ \Delta > U_d$,  one can further simplify to $K \sim \frac{3}{4} \frac{\lambda_p^2 t_{pd\sigma}^4}{U_d \Delta^4} \equiv \frac{3}{4} \frac{t_{\rm ind}^2}{U_d}$, where $t_{\rm ind} = \frac{\lambda_p t_{pd\sigma}^2}{\Delta^2}$ describes
the effective hopping between the M and M site via the A sites.
When the cubic symmetry
 is slightly broken, 
a slight difference between $K^z$ and $K^{x/y}$ appears as shown in the SM.
The Heisenberg interaction $J_{\rm ind}$ via the superexchange process is ferromagnetic and its strength is half of $K$ term.
Interestingly, $J_{\rm ind}$ is finite when the large SOC of the anion sites is present, even when Hund's coupling $J_{H_p}$ is absent.
For the other limit of $\lambda_p \rightarrow 0$, the ferromagnetic Heisenberg interaction from two-hole processes is found and
the Kitaev term vanishes. 

There is also a direct hopping $t$ between the M sites, which leads to the antiferromagnetic Heisenberg term $J_{\rm dir} \sim 4 t^2/U_d$. 
Given the distance between M and A sites vs. M sites, the direct hopping integral $t$ is an order of magnitude smaller than the indirect $t_1,\ t_2,\ t_3$ hoppings, however,
the perturbation process involves second order terms.
Thus the antiferromagnetic Heisenberg term $J_{\rm dir}$ of similar strength to the ferromagnetic term $J_{\rm ind}$ may be generated via direct hopping. 
Since the direct and indirect Heisenberg terms come with opposite signs,
one may expect 
a small total Heisenberg interaction  $J \equiv J_{\rm dir} - |J_{\rm ind}|$ and the Kitaev interaction dominates the physics of the spin S=1 systems. 

\begin{figure}[!ht]
  \centering
\begin{overpic}[width=1.0\linewidth]{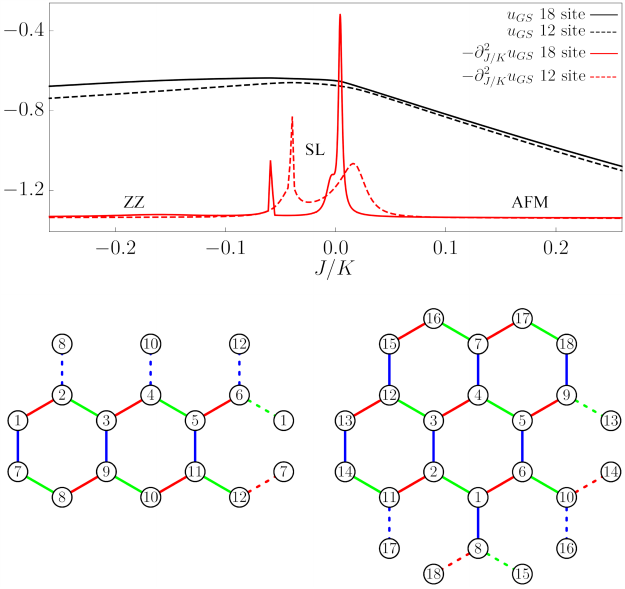}
\put(0,92){(a)}
\put(0,44){(b)}
\put(50,44){(c)}
\end{overpic}
  \caption{ (a) The phase diagram of the S=1 KH model. By tuning the ratio of J/K, two transitions
  signalled by the singular behavior of second derivative of the ground state energy density $u_{gs}$
  are found on both 12- and 18-site ED clusters shown in (b) and (c), respectively. Energy density units are $\sqrt{J^2+K^2}/N$. 
  There are three phases identified by spin-spin correlators as discussed in the main text.
   The Kitaev spin liquid (SL) appears near $J/K \sim 0$,
  and AFM and ZZ orderings are respectively found in the antiferromagnetic and ferromagnetic Heisenberg interaction regions. }
  \label{fig:edresults}
\end{figure}

\textit{Exact Diagonalization of S=1 KH model --}
We show that the n.n. spin-model of two-electrons in the $e_g$ orbitals surrounded by anions with strong SOC forming edge-shared octahedra
consists of the S=1 Kitaev and Heisenberg interactions. 
 It is worthwhile to check if the S=1 Kitaev spin liquid survives in the presence of the Heisenberg term. 
 We carry out ED calculations to determine the phase diagram near the antiferromagnetic Kitaev term.
The ED results are shown in Fig. \ref{fig:edresults} (a) on two clusters of N=12 and N=18 sites
using the periodic boundary conditions in Fig.\ref{fig:edresults} (b) and (c), respectively. 
Phase transitions are identified by the singular behavior of the second derivative of the ground state energy density ($u_{GS}$) with respect to the variable $J/K$, i.e., $-\partial^2_{J/K} u_{GS}$.

Our results show three phases. A finite region of the Kitaev phase around the antiferromagnetic $K$ region appears for both clusters. 
To clarify the nature of the phases, we examine the spin-spin correlation of the three regions. The Kitaev phase has a finite n.n. correlation
and further neighbor correlations are zero, consistent with the pure Kitaev S=1 phase.\cite{Baskaran2008PRB}. 
For $J/K \sim 0.3$ we find the antiferromagnetic (AFM) ordered phase, while for $J/K \sim -0.3$, the zig-zag (ZZ) ordered phase is found. 
These magnetically ordered phases match the magnetic orderings found in the S=$\frac{1}{2}$ case\cite{ChaloupkaPRL2013}.

\textit{Kitaev candiate materials } --  
A single layer of NiI$_2$ is a candidate for S=1 Kitaev materials on triangular lattice.
The triangular lattice has X-,Y-, and Z-bond defined similarly to the honeycomb lattice and the above derivation of the mechanism is applicable.
The bulk compounds form triangular layers of Ni cations and I anions form edge-shared octahedral cages around Ni.  
 While the bulk NiCl$_2$ is ferromagnetic below 52 K \cite{BuseyJoACS1952}, the heavier sister compound NiI$_2$ has helimagnetic order below 75 K \cite{Billerey1977PLA,KuindersmaPhysica1981}.
 The helical ordering in the bulk compound is related to the layer coupling as the ordering wave vector involves the lattice vector perpendicular to
 the triangular layer.\cite{KuindersmaPhysica1981,McGuire2015CoM} Thus a single layer of NiI$_2$ is desirable to test the dominant Kitaev interaction.

 Another group of potential materials is 
the layered transitional metal (M) oxide compounds A$_3$Ni$_2$XO$_6$ (A=Li, Na, X=Bi, Sb).
Unlike the simple binary NiI$_2$, the M sites are surrounded by edge-shared oxygen octahedral cages, forming layers of honeycomb networks sandwiched between layers of the alkali A sites. X sites reside in the center of the honeycomb.
A$_3$Ni$_2$XO$_6$ exhibits ZZ ordering at low temperatures.\cite{ZverevaPRB2015,NalbandyanPRB2017}
There are 2 electrons in e$_g$ orbitals making total spin S=1 states, a good example for the proposed mechanism.
The strong SOC may occur via proximity to the heavy X atoms.
While the oxygen has a weak atomic SOC,  
the heavy X atoms with strong SOC $\lambda_{\rm X}$ induce splitting among the p orbitals of the oxygen atoms
 leading to similar effects presented above.
For instance, the effective SOC could be enhanced when one considers hopping between X and O sites denoted by $t_{pp}$.
Using a perturbative approach, the strength of effective SOC in the p orbitals of O sites is then determined by ${\tilde \lambda}_p \sim \left(\frac{t_{pp}^2}{{\tilde \Delta} - \frac{\lambda_{\rm X}}{2}} -
\frac{t_{pp}^2}{{\tilde \Delta}  + \lambda_{\rm X}}\right) $ where ${\tilde \Delta}$ is an atomic potential difference between X and O atoms.
While it is difficult to quantify ${\tilde \lambda}_p$ in this case, we note that the specific heat measurements resulting in entropy of $\sim \frac{1}{2} {\rm log 3}$ per Ni above the Neel temperature for both Li$_3$Ni$_2$SbO$_6$ and Na$_3$Ni$_2$SbO$_6$ \cite{ZverevaPRB2015} strongly hint that they are promising candidates
for S=1 Kitaev honeycomb materials. 

\textit{ Outlook and Summary ---}
The bond-dependent interactions are ubiquitous in Mott insulators with edge-shared octahedral environment and strong SOC.
This is because SOC mixes different orbitals and spin components at a given site, and bond-dependent spin interactions
rely on the hopping integrals of the bond, whose size is determined from the overlap of relevant orbitals.
In an edge-shared environment, bond-dependent terms could be dominant over the conventional Heisenberg terms. 
Despite its ubiquitousness, only candidates of spin $\frac{1}{2}$ have been investigated so far,
because the mechanism of such interactions has not been explored for higher spins. 

Here we  derive a microscopic S=1 Kitaev interaction via superexchange processes between
 half-filled e$_g$ orbital cations mediated by p orbital anions with strong SOC using the standard strong coupling expansion. 
We find the dominant interaction is the antiferromagnetic Kitaev term, whose strength is twice as big as that of the ferromagnetic Heisenberg interaction.
Taking into account the direct exchange process that results in an antiferromagnetic Heisenberg term,
 we expect that the Kitaev interaction dominates spin physics of 
 these Mott insulators. 
A small region of S=1 Kitaev phase with only n.n. spin-spin correlation is found in 12- and 18-site ED calculations. 
A finite ferromagnetic Heisenberg interaction stabilizes the ZZ magnetic ordering nearby the spin liquid.
S=1 Kitaev candidates include a single layer of NiI$_2$ on the triangular lattice
and A$_3$Ni$_2$XO$_6$ with X=Bi, Sb and A=Li, Na on the honeycomb lattice.  

The analysis presented in the current work can be extended to a higher spin Kitaev model.
For example, Cr$^{3+}$ leaves three electrons in the t$_{2g}$ orbitals that make spin $\frac{3}{2}$ via Hund's coupling and
the superexchange processes via strong SOC anions lead to the Kitaev term.  
Thus CrI$_3$ is a candidate for the spin S=$\frac{3}{2}$ Kitaev Mott insulator.
A single layer of CrI$_3$ shows a ferromagnetic ordering with strong anisotropy\cite{Dillon1965JoAP,McGuire2015CoM,Huang2017Nature}
 and investigating the microscopic mechanism of such anisotropy from the bond-dependent interactions is an excellent future study. 

A group of van der Waals transition metal halides, such as MX$_2$ and MX$_3$, provides a rich family of magnetic materials.
The dihalides MX$_2$ and trihalides MX$_3$ are made of triangular and honeycomb networks of transition metal cations respectively, surrounded by edge-shared anions X.\cite{McGuire2017Crystals}
When X is heavy,
the strong SOC at X sites plays a role in the magnetic mechanism presented in this work.
  Theoretical studies on these magnetic materials have been limited to the first, second, and third n.n. Heisenberg model.
We propose to revisit these layered 3d transitional metal compounds with edge-shared heavy anions
from a new perspective of bond-dependent interaction.
 
There are various experimental ways to test the Kitaev interactions in these candidate materials. 
Inelastic neutron scattering measurement allows to map the microscopic spin interactions in these Mott insulator. 
The magnetic field is a way to induce or reveal the Kitaev spin liquids and its effects have been widely studied in $\alpha$-RuCl$_3$.  
\cite{Yadav2016SCIR,Baek2017PRL,Wolter2017PRB,Zheng2017PRL,
  Jansa2018NAP,Fu2018PRB,GohlkePRB2018,NasuPRB2018,
  Ronquillo2018ARXIV,
 Ciaran2018NC,LiangPRB2018,Lampenkelley2018ARXIV,
 Lu2018ARXIV,Zou2018ARXIV,Patel2018ARXIV,Jacob2019ARXIV}
Note that the S=1 Kitaev materials suggested here have the antiferromagnetic Kitaev interaction dominant, 
unlike the S=$\frac{1}{2}$ Kitaev candidate RuCl$_3$
that has the ferromagnetic Kitaev interaction dominant. 
Thus the magnetic field along [111] direction may induce the U(1) spin liquid with Fermi surface,
similar to S=$\frac{1}{2}$ case\cite{Ciaran2018NC,Fu2018PRB,Ronquillo2018ARXIV,Lu2018ARXIV,Zou2018ARXIV,Patel2018ARXIV}. 
Theoretical studies on S=1 Kitaev materials  and experimental studies on a single layer of the proposed materials with and without the magnetic field
are interesting projects to pursue in the future.
Determining S=$\frac{3}{2}$ Kitaev materials and their magnetic field dependence are also excellent tasks for future studies.

[Note added] While preparing the manuscript, we note Ref. \onlinecite{CrI3arxiv2019} where a ferromagnetic S=$\frac{3}{2}$ Kitaev interaction is suggested to
understand the angle-dependent ferromagnetic resonance experimental data on CrI$_3$. 

\textit{Acknowlegment} We thank A. Catuneanu for useful discussions during the early stage of the project. 
This work was supported by the Natural Sciences and Engineering Research Council of Canada and the Center for Quantum Materials at the University of Toronto. Computations were performed on the Niagara supercomputer at the SciNet HPC Consortium. SciNet is funded by: the Canada Foundation for Innovation under the auspices of Compute Canada; the Government of Ontario; Ontario Research Fund - Research Excellence; and the University of Toronto.

\bibliography{kitaev_s1_model}

\begin{thebibliography}{60}%
\makeatletter
\providecommand \@ifxundefined [1]{%
 \@ifx{#1\undefined}
}%
\providecommand \@ifnum [1]{%
 \ifnum #1\expandafter \@firstoftwo
 \else \expandafter \@secondoftwo
 \fi
}%
\providecommand \@ifx [1]{%
 \ifx #1\expandafter \@firstoftwo
 \else \expandafter \@secondoftwo
 \fi
}%
\providecommand \natexlab [1]{#1}%
\providecommand \enquote  [1]{``#1''}%
\providecommand \bibnamefont  [1]{#1}%
\providecommand \bibfnamefont [1]{#1}%
\providecommand \citenamefont [1]{#1}%
\providecommand \href@noop [0]{\@secondoftwo}%
\providecommand \href [0]{\begingroup \@sanitize@url \@href}%
\providecommand \@href[1]{\@@startlink{#1}\@@href}%
\providecommand \@@href[1]{\endgroup#1\@@endlink}%
\providecommand \@sanitize@url [0]{\catcode `\\12\catcode `\$12\catcode
  `\&12\catcode `\#12\catcode `\^12\catcode `\_12\catcode `\%12\relax}%
\providecommand \@@startlink[1]{}%
\providecommand \@@endlink[0]{}%
\providecommand \url  [0]{\begingroup\@sanitize@url \@url }%
\providecommand \@url [1]{\endgroup\@href {#1}{\urlprefix }}%
\providecommand \urlprefix  [0]{URL }%
\providecommand \Eprint [0]{\href }%
\providecommand \doibase [0]{http://dx.doi.org/}%
\providecommand \selectlanguage [0]{\@gobble}%
\providecommand \bibinfo  [0]{\@secondoftwo}%
\providecommand \bibfield  [0]{\@secondoftwo}%
\providecommand \translation [1]{[#1]}%
\providecommand \BibitemOpen [0]{}%
\providecommand \bibitemStop [0]{}%
\providecommand \bibitemNoStop [0]{.\EOS\space}%
\providecommand \EOS [0]{\spacefactor3000\relax}%
\providecommand \BibitemShut  [1]{\csname bibitem#1\endcsname}%
\let\auto@bib@innerbib\@empty
\bibitem [{\citenamefont {Balents}(2010)}]{Balents2010NA}%
  \BibitemOpen
  \bibfield  {author} {\bibinfo {author} {\bibfnamefont {L.}~\bibnamefont
  {Balents}},\ }\href {http://dx.doi.org/10.1038/nature08917} {\bibfield
  {journal} {\bibinfo  {journal} {Nature}\ }\textbf {\bibinfo {volume} {464}},\
  \bibinfo {pages} {199} (\bibinfo {year} {2010})}\BibitemShut {NoStop}%
\bibitem [{\citenamefont {Zhou}\ \emph {et~al.}(2017)\citenamefont {Zhou},
  \citenamefont {Kanoda},\ and\ \citenamefont {Ng}}]{Zhou2017RMP}%
  \BibitemOpen
  \bibfield  {author} {\bibinfo {author} {\bibfnamefont {Y.}~\bibnamefont
  {Zhou}}, \bibinfo {author} {\bibfnamefont {K.}~\bibnamefont {Kanoda}}, \ and\
  \bibinfo {author} {\bibfnamefont {T.-K.}\ \bibnamefont {Ng}},\ }\href
  {\doibase 10.1103/RevModPhys.89.025003} {\bibfield  {journal} {\bibinfo
  {journal} {Rev. Mod. Phys.}\ }\textbf {\bibinfo {volume} {89}},\ \bibinfo
  {pages} {025003} (\bibinfo {year} {2017})}\BibitemShut {NoStop}%
\bibitem [{\citenamefont {Kitaev}(2006)}]{Kitaev2006AoP}%
  \BibitemOpen
  \bibfield  {author} {\bibinfo {author} {\bibfnamefont {A.}~\bibnamefont
  {Kitaev}},\ }\href
  {http://www.sciencedirect.com/science/article/pii/S0003491605002381"}
  {\bibfield  {journal} {\bibinfo  {journal} {Annals of Physics}\ }\textbf
  {\bibinfo {volume} {321}},\ \bibinfo {pages} {2 } (\bibinfo {year}
  {2006})}\BibitemShut {NoStop}%
\bibitem [{\citenamefont {Kasahara}\ \emph {et~al.}(2018)\citenamefont
  {Kasahara}, \citenamefont {Ohnishi}, \citenamefont {Mizukami}, \citenamefont
  {Tanaka}, \citenamefont {Ma}, \citenamefont {Sugii}, \citenamefont {Kurita},
  \citenamefont {Tanaka}, \citenamefont {Nasu}, \citenamefont {Motome},
  \citenamefont {Shibauchi},\ and\ \citenamefont {Matsuda}}]{Matsuda2018NA}%
  \BibitemOpen
  \bibfield  {author} {\bibinfo {author} {\bibfnamefont {Y.}~\bibnamefont
  {Kasahara}}, \bibinfo {author} {\bibfnamefont {T.}~\bibnamefont {Ohnishi}},
  \bibinfo {author} {\bibfnamefont {Y.}~\bibnamefont {Mizukami}}, \bibinfo
  {author} {\bibfnamefont {O.}~\bibnamefont {Tanaka}}, \bibinfo {author}
  {\bibfnamefont {S.}~\bibnamefont {Ma}}, \bibinfo {author} {\bibfnamefont
  {K.}~\bibnamefont {Sugii}}, \bibinfo {author} {\bibfnamefont
  {N.}~\bibnamefont {Kurita}}, \bibinfo {author} {\bibfnamefont
  {H.}~\bibnamefont {Tanaka}}, \bibinfo {author} {\bibfnamefont
  {J.}~\bibnamefont {Nasu}}, \bibinfo {author} {\bibfnamefont {Y.}~\bibnamefont
  {Motome}}, \bibinfo {author} {\bibfnamefont {T.}~\bibnamefont {Shibauchi}}, \
  and\ \bibinfo {author} {\bibfnamefont {Y.}~\bibnamefont {Matsuda}},\ }\href
  {\doibase 10.1038/s41586-018-0274-0} {\bibfield  {journal} {\bibinfo
  {journal} {Nature}\ }\textbf {\bibinfo {volume} {559}},\ \bibinfo {pages}
  {227} (\bibinfo {year} {2018})}\BibitemShut {NoStop}%
\bibitem [{\citenamefont {Jackeli}\ and\ \citenamefont
  {Khaliullin}(2009)}]{Khaliullin2009PRL}%
  \BibitemOpen
  \bibfield  {author} {\bibinfo {author} {\bibfnamefont {G.}~\bibnamefont
  {Jackeli}}\ and\ \bibinfo {author} {\bibfnamefont {G.}~\bibnamefont
  {Khaliullin}},\ }\href {\doibase 10.1103/PhysRevLett.102.017205} {\bibfield
  {journal} {\bibinfo  {journal} {Phys. Rev. Lett.}\ }\textbf {\bibinfo
  {volume} {102}},\ \bibinfo {pages} {017205} (\bibinfo {year}
  {2009})}\BibitemShut {NoStop}%
\bibitem [{\citenamefont {Chaloupka}\ \emph {et~al.}(2010)\citenamefont
  {Chaloupka}, \citenamefont {Jackeli},\ and\ \citenamefont
  {Khaliullin}}]{Chaloupka2010PRL}%
  \BibitemOpen
  \bibfield  {author} {\bibinfo {author} {\bibfnamefont {J.}~\bibnamefont
  {Chaloupka}}, \bibinfo {author} {\bibfnamefont {G.}~\bibnamefont {Jackeli}},
  \ and\ \bibinfo {author} {\bibfnamefont {G.}~\bibnamefont {Khaliullin}},\
  }\href {\doibase 10.1103/PhysRevLett.105.027204} {\bibfield  {journal}
  {\bibinfo  {journal} {Phys. Rev. Lett.}\ }\textbf {\bibinfo {volume} {105}},\
  \bibinfo {pages} {027204} (\bibinfo {year} {2010})}\BibitemShut {NoStop}%
\bibitem [{\citenamefont {Singh}\ \emph {et~al.}(2012)\citenamefont {Singh},
  \citenamefont {Manni}, \citenamefont {Reuther}, \citenamefont {Berlijn},
  \citenamefont {Thomale}, \citenamefont {Ku}, \citenamefont {Trebst},\ and\
  \citenamefont {Gegenwart}}]{Gegenwart2012PRL}%
  \BibitemOpen
  \bibfield  {author} {\bibinfo {author} {\bibfnamefont {Y.}~\bibnamefont
  {Singh}}, \bibinfo {author} {\bibfnamefont {S.}~\bibnamefont {Manni}},
  \bibinfo {author} {\bibfnamefont {J.}~\bibnamefont {Reuther}}, \bibinfo
  {author} {\bibfnamefont {T.}~\bibnamefont {Berlijn}}, \bibinfo {author}
  {\bibfnamefont {R.}~\bibnamefont {Thomale}}, \bibinfo {author} {\bibfnamefont
  {W.}~\bibnamefont {Ku}}, \bibinfo {author} {\bibfnamefont {S.}~\bibnamefont
  {Trebst}}, \ and\ \bibinfo {author} {\bibfnamefont {P.}~\bibnamefont
  {Gegenwart}},\ }\href {\doibase 10.1103/PhysRevLett.108.127203} {\bibfield
  {journal} {\bibinfo  {journal} {Phys. Rev. Lett.}\ }\textbf {\bibinfo
  {volume} {108}},\ \bibinfo {pages} {127203} (\bibinfo {year}
  {2012})}\BibitemShut {NoStop}%
\bibitem [{\citenamefont {Choi}\ \emph {et~al.}(2012)\citenamefont {Choi},
  \citenamefont {Coldea}, \citenamefont {Kolmogorov}, \citenamefont
  {Lancaster}, \citenamefont {Mazin}, \citenamefont {Blundell}, \citenamefont
  {Radaelli}, \citenamefont {Singh}, \citenamefont {Gegenwart}, \citenamefont
  {Choi}, \citenamefont {Cheong}, \citenamefont {Baker}, \citenamefont
  {Stock},\ and\ \citenamefont {Taylor}}]{Taylor2012PRL}%
  \BibitemOpen
  \bibfield  {author} {\bibinfo {author} {\bibfnamefont {S.~K.}\ \bibnamefont
  {Choi}}, \bibinfo {author} {\bibfnamefont {R.}~\bibnamefont {Coldea}},
  \bibinfo {author} {\bibfnamefont {A.~N.}\ \bibnamefont {Kolmogorov}},
  \bibinfo {author} {\bibfnamefont {T.}~\bibnamefont {Lancaster}}, \bibinfo
  {author} {\bibfnamefont {I.~I.}\ \bibnamefont {Mazin}}, \bibinfo {author}
  {\bibfnamefont {S.~J.}\ \bibnamefont {Blundell}}, \bibinfo {author}
  {\bibfnamefont {P.~G.}\ \bibnamefont {Radaelli}}, \bibinfo {author}
  {\bibfnamefont {Y.}~\bibnamefont {Singh}}, \bibinfo {author} {\bibfnamefont
  {P.}~\bibnamefont {Gegenwart}}, \bibinfo {author} {\bibfnamefont {K.~R.}\
  \bibnamefont {Choi}}, \bibinfo {author} {\bibfnamefont {S.-W.}\ \bibnamefont
  {Cheong}}, \bibinfo {author} {\bibfnamefont {P.~J.}\ \bibnamefont {Baker}},
  \bibinfo {author} {\bibfnamefont {C.}~\bibnamefont {Stock}}, \ and\ \bibinfo
  {author} {\bibfnamefont {J.}~\bibnamefont {Taylor}},\ }\href {\doibase
  10.1103/PhysRevLett.108.127204} {\bibfield  {journal} {\bibinfo  {journal}
  {Phys. Rev. Lett.}\ }\textbf {\bibinfo {volume} {108}},\ \bibinfo {pages}
  {127204} (\bibinfo {year} {2012})}\BibitemShut {NoStop}%
\bibitem [{\citenamefont {Chaloupka}\ \emph {et~al.}(2013)\citenamefont
  {Chaloupka}, \citenamefont {Jackeli},\ and\ \citenamefont
  {Khaliullin}}]{ChaloupkaPRL2013}%
  \BibitemOpen
  \bibfield  {author} {\bibinfo {author} {\bibfnamefont {J.}~\bibnamefont
  {Chaloupka}}, \bibinfo {author} {\bibfnamefont {G.}~\bibnamefont {Jackeli}},
  \ and\ \bibinfo {author} {\bibfnamefont {G.}~\bibnamefont {Khaliullin}},\
  }\href {\doibase 10.1103/PhysRevLett.110.097204} {\bibfield  {journal}
  {\bibinfo  {journal} {Phys. Rev. Lett.}\ }\textbf {\bibinfo {volume} {110}},\
  \bibinfo {pages} {097204} (\bibinfo {year} {2013})}\BibitemShut {NoStop}%
\bibitem [{\citenamefont {Witczak-Krempa}\ \emph {et~al.}(2014)\citenamefont
  {Witczak-Krempa}, \citenamefont {Chen}, \citenamefont {Kim},\ and\
  \citenamefont {Balents}}]{Balents2014ARoCMP}%
  \BibitemOpen
  \bibfield  {author} {\bibinfo {author} {\bibfnamefont {W.}~\bibnamefont
  {Witczak-Krempa}}, \bibinfo {author} {\bibfnamefont {G.}~\bibnamefont
  {Chen}}, \bibinfo {author} {\bibfnamefont {Y.~B.}\ \bibnamefont {Kim}}, \
  and\ \bibinfo {author} {\bibfnamefont {L.}~\bibnamefont {Balents}},\ }\href
  {\doibase 10.1146/annurev-conmatphys-020911-125138} {\bibfield  {journal}
  {\bibinfo  {journal} {Annual Review of Condensed Matter Physics}\ }\textbf
  {\bibinfo {volume} {5}},\ \bibinfo {pages} {57} (\bibinfo {year}
  {2014})}\BibitemShut {NoStop}%
\bibitem [{\citenamefont {Rau}\ \emph {et~al.}(2014)\citenamefont {Rau},
  \citenamefont {Lee},\ and\ \citenamefont {Kee}}]{Rau2014PRL}%
  \BibitemOpen
  \bibfield  {author} {\bibinfo {author} {\bibfnamefont {J.~G.}\ \bibnamefont
  {Rau}}, \bibinfo {author} {\bibfnamefont {E.~K.-H.}\ \bibnamefont {Lee}}, \
  and\ \bibinfo {author} {\bibfnamefont {H.-Y.}\ \bibnamefont {Kee}},\ }\href
  {\doibase 10.1103/PhysRevLett.112.077204} {\bibfield  {journal} {\bibinfo
  {journal} {Phys. Rev. Lett.}\ }\textbf {\bibinfo {volume} {112}},\ \bibinfo
  {pages} {077204} (\bibinfo {year} {2014})}\BibitemShut {NoStop}%
\bibitem [{\citenamefont {Plumb}\ \emph {et~al.}(2014)\citenamefont {Plumb},
  \citenamefont {Clancy}, \citenamefont {Sandilands}, \citenamefont {Shankar},
  \citenamefont {Hu}, \citenamefont {Burch}, \citenamefont {Kee},\ and\
  \citenamefont {Kim}}]{YJKim2014PRB}%
  \BibitemOpen
  \bibfield  {author} {\bibinfo {author} {\bibfnamefont {K.~W.}\ \bibnamefont
  {Plumb}}, \bibinfo {author} {\bibfnamefont {J.~P.}\ \bibnamefont {Clancy}},
  \bibinfo {author} {\bibfnamefont {L.~J.}\ \bibnamefont {Sandilands}},
  \bibinfo {author} {\bibfnamefont {V.~V.}\ \bibnamefont {Shankar}}, \bibinfo
  {author} {\bibfnamefont {Y.~F.}\ \bibnamefont {Hu}}, \bibinfo {author}
  {\bibfnamefont {K.~S.}\ \bibnamefont {Burch}}, \bibinfo {author}
  {\bibfnamefont {H.-Y.}\ \bibnamefont {Kee}}, \ and\ \bibinfo {author}
  {\bibfnamefont {Y.-J.}\ \bibnamefont {Kim}},\ }\href {\doibase
  10.1103/PhysRevB.90.041112} {\bibfield  {journal} {\bibinfo  {journal} {Phys.
  Rev. B}\ }\textbf {\bibinfo {volume} {90}},\ \bibinfo {pages} {041112}
  (\bibinfo {year} {2014})}\BibitemShut {NoStop}%
\bibitem [{\citenamefont {Rau}\ and\ \citenamefont {Kee}()}]{Rau2014ARXIV}%
  \BibitemOpen
  \bibfield  {author} {\bibinfo {author} {\bibfnamefont {J.~G.}\ \bibnamefont
  {Rau}}\ and\ \bibinfo {author} {\bibfnamefont {H.-Y.}\ \bibnamefont {Kee}},\
  }\href@noop {} {\ }\Eprint {http://arxiv.org/abs/arXiv:1408.4811}
  {arXiv:1408.4811} \BibitemShut {NoStop}%
\bibitem [{\citenamefont {Sears}\ \emph {et~al.}(2015)\citenamefont {Sears},
  \citenamefont {Songvilay}, \citenamefont {Plumb}, \citenamefont {Clancy},
  \citenamefont {Qiu}, \citenamefont {Zhao}, \citenamefont {Parshall},\ and\
  \citenamefont {Kim}}]{Sears2015PRB}%
  \BibitemOpen
  \bibfield  {author} {\bibinfo {author} {\bibfnamefont {J.~A.}\ \bibnamefont
  {Sears}}, \bibinfo {author} {\bibfnamefont {M.}~\bibnamefont {Songvilay}},
  \bibinfo {author} {\bibfnamefont {K.~W.}\ \bibnamefont {Plumb}}, \bibinfo
  {author} {\bibfnamefont {J.~P.}\ \bibnamefont {Clancy}}, \bibinfo {author}
  {\bibfnamefont {Y.}~\bibnamefont {Qiu}}, \bibinfo {author} {\bibfnamefont
  {Y.}~\bibnamefont {Zhao}}, \bibinfo {author} {\bibfnamefont {D.}~\bibnamefont
  {Parshall}}, \ and\ \bibinfo {author} {\bibfnamefont {Y.-J.}\ \bibnamefont
  {Kim}},\ }\href {\doibase 10.1103/PhysRevB.91.144420} {\bibfield  {journal}
  {\bibinfo  {journal} {Phys. Rev. B}\ }\textbf {\bibinfo {volume} {91}},\
  \bibinfo {pages} {144420} (\bibinfo {year} {2015})}\BibitemShut {NoStop}%
\bibitem [{\citenamefont {Kim}\ \emph {et~al.}(2015)\citenamefont {Kim},
  \citenamefont {V.}, \citenamefont {Catuneanu},\ and\ \citenamefont
  {Kee}}]{HSKim2015PRB}%
  \BibitemOpen
  \bibfield  {author} {\bibinfo {author} {\bibfnamefont {H.-S.}\ \bibnamefont
  {Kim}}, \bibinfo {author} {\bibfnamefont {V.~S.}\ \bibnamefont {V.}},
  \bibinfo {author} {\bibfnamefont {A.}~\bibnamefont {Catuneanu}}, \ and\
  \bibinfo {author} {\bibfnamefont {H.-Y.}\ \bibnamefont {Kee}},\ }\href
  {\doibase 10.1103/PhysRevB.91.241110} {\bibfield  {journal} {\bibinfo
  {journal} {Phys. Rev. B}\ }\textbf {\bibinfo {volume} {91}},\ \bibinfo
  {pages} {241110} (\bibinfo {year} {2015})}\BibitemShut {NoStop}%
\bibitem [{\citenamefont {Sandilands}\ \emph {et~al.}(2016)\citenamefont
  {Sandilands}, \citenamefont {Tian}, \citenamefont {Reijnders}, \citenamefont
  {Kim}, \citenamefont {Plumb}, \citenamefont {Kim}, \citenamefont {Kee},\ and\
  \citenamefont {Burch}}]{Sandilands2016PRB}%
  \BibitemOpen
  \bibfield  {author} {\bibinfo {author} {\bibfnamefont {L.~J.}\ \bibnamefont
  {Sandilands}}, \bibinfo {author} {\bibfnamefont {Y.}~\bibnamefont {Tian}},
  \bibinfo {author} {\bibfnamefont {A.~A.}\ \bibnamefont {Reijnders}}, \bibinfo
  {author} {\bibfnamefont {H.-S.}\ \bibnamefont {Kim}}, \bibinfo {author}
  {\bibfnamefont {K.~W.}\ \bibnamefont {Plumb}}, \bibinfo {author}
  {\bibfnamefont {Y.-J.}\ \bibnamefont {Kim}}, \bibinfo {author} {\bibfnamefont
  {H.-Y.}\ \bibnamefont {Kee}}, \ and\ \bibinfo {author} {\bibfnamefont
  {K.~S.}\ \bibnamefont {Burch}},\ }\href {\doibase 10.1103/PhysRevB.93.075144}
  {\bibfield  {journal} {\bibinfo  {journal} {Phys. Rev. B}\ }\textbf {\bibinfo
  {volume} {93}},\ \bibinfo {pages} {075144} (\bibinfo {year}
  {2016})}\BibitemShut {NoStop}%
\bibitem [{\citenamefont {Sandilands}\ \emph {et~al.}(2015)\citenamefont
  {Sandilands}, \citenamefont {Tian}, \citenamefont {Plumb}, \citenamefont
  {Kim},\ and\ \citenamefont {Burch}}]{Sandilands2015PRL}%
  \BibitemOpen
  \bibfield  {author} {\bibinfo {author} {\bibfnamefont {L.~J.}\ \bibnamefont
  {Sandilands}}, \bibinfo {author} {\bibfnamefont {Y.}~\bibnamefont {Tian}},
  \bibinfo {author} {\bibfnamefont {K.~W.}\ \bibnamefont {Plumb}}, \bibinfo
  {author} {\bibfnamefont {Y.-J.}\ \bibnamefont {Kim}}, \ and\ \bibinfo
  {author} {\bibfnamefont {K.~S.}\ \bibnamefont {Burch}},\ }\href {\doibase
  10.1103/PhysRevLett.114.147201} {\bibfield  {journal} {\bibinfo  {journal}
  {Phys. Rev. Lett.}\ }\textbf {\bibinfo {volume} {114}},\ \bibinfo {pages}
  {147201} (\bibinfo {year} {2015})}\BibitemShut {NoStop}%
\bibitem [{\citenamefont {Banerjee}\ \emph {et~al.}(2016)\citenamefont
  {Banerjee}, \citenamefont {Bridges}, \citenamefont {Yan}, \citenamefont
  {Aczel}, \citenamefont {Li}, \citenamefont {Stone}, \citenamefont {Granroth},
  \citenamefont {Lumsden}, \citenamefont {Yiu}, \citenamefont {Knolle},
  \citenamefont {Bhattacharjee}, \citenamefont {Kovrizhin}, \citenamefont
  {Moessner}, \citenamefont {Tennant}, \citenamefont {Mandrus},\ and\
  \citenamefont {Nagler}}]{Banerjee2016NAM}%
  \BibitemOpen
  \bibfield  {author} {\bibinfo {author} {\bibfnamefont {A.}~\bibnamefont
  {Banerjee}}, \bibinfo {author} {\bibfnamefont {C.~A.}\ \bibnamefont
  {Bridges}}, \bibinfo {author} {\bibfnamefont {J.-Q.}\ \bibnamefont {Yan}},
  \bibinfo {author} {\bibfnamefont {A.~A.}\ \bibnamefont {Aczel}}, \bibinfo
  {author} {\bibfnamefont {L.}~\bibnamefont {Li}}, \bibinfo {author}
  {\bibfnamefont {M.~B.}\ \bibnamefont {Stone}}, \bibinfo {author}
  {\bibfnamefont {G.~E.}\ \bibnamefont {Granroth}}, \bibinfo {author}
  {\bibfnamefont {M.~D.}\ \bibnamefont {Lumsden}}, \bibinfo {author}
  {\bibfnamefont {Y.}~\bibnamefont {Yiu}}, \bibinfo {author} {\bibfnamefont
  {J.}~\bibnamefont {Knolle}}, \bibinfo {author} {\bibfnamefont
  {S.}~\bibnamefont {Bhattacharjee}}, \bibinfo {author} {\bibfnamefont {D.~L.}\
  \bibnamefont {Kovrizhin}}, \bibinfo {author} {\bibfnamefont {R.}~\bibnamefont
  {Moessner}}, \bibinfo {author} {\bibfnamefont {D.~A.}\ \bibnamefont
  {Tennant}}, \bibinfo {author} {\bibfnamefont {D.~G.}\ \bibnamefont
  {Mandrus}}, \ and\ \bibinfo {author} {\bibfnamefont {S.~E.}\ \bibnamefont
  {Nagler}},\ }\href {http://dx.doi.org/10.1038/nmat4604} {\bibfield  {journal}
  {\bibinfo  {journal} {Nature Materials}\ }\textbf {\bibinfo {volume} {15}},\
  \bibinfo {pages} {733} (\bibinfo {year} {2016})}\BibitemShut {NoStop}%
\bibitem [{\citenamefont {Rau}\ \emph {et~al.}(2016)\citenamefont {Rau},
  \citenamefont {Lee},\ and\ \citenamefont {Kee}}]{Rau2016ARoCMP}%
  \BibitemOpen
  \bibfield  {author} {\bibinfo {author} {\bibfnamefont {J.~G.}\ \bibnamefont
  {Rau}}, \bibinfo {author} {\bibfnamefont {E.~K.-H.}\ \bibnamefont {Lee}}, \
  and\ \bibinfo {author} {\bibfnamefont {H.-Y.}\ \bibnamefont {Kee}},\ }\href
  {\doibase 10.1146/annurev-conmatphys-031115-011319} {\bibfield  {journal}
  {\bibinfo  {journal} {Annual Review of Condensed Matter Physics}\ }\textbf
  {\bibinfo {volume} {7}},\ \bibinfo {pages} {195} (\bibinfo {year}
  {2016})}\BibitemShut {NoStop}%
\bibitem [{\citenamefont {Johnson}\ \emph {et~al.}(2015)\citenamefont
  {Johnson}, \citenamefont {Williams}, \citenamefont {Haghighirad},
  \citenamefont {Singleton}, \citenamefont {Zapf}, \citenamefont {Manuel},
  \citenamefont {Mazin}, \citenamefont {Li}, \citenamefont {Jeschke},
  \citenamefont {Valent\'{\i}},\ and\ \citenamefont {Coldea}}]{Johnson2015PRB}%
  \BibitemOpen
  \bibfield  {author} {\bibinfo {author} {\bibfnamefont {R.~D.}\ \bibnamefont
  {Johnson}}, \bibinfo {author} {\bibfnamefont {S.~C.}\ \bibnamefont
  {Williams}}, \bibinfo {author} {\bibfnamefont {A.~A.}\ \bibnamefont
  {Haghighirad}}, \bibinfo {author} {\bibfnamefont {J.}~\bibnamefont
  {Singleton}}, \bibinfo {author} {\bibfnamefont {V.}~\bibnamefont {Zapf}},
  \bibinfo {author} {\bibfnamefont {P.}~\bibnamefont {Manuel}}, \bibinfo
  {author} {\bibfnamefont {I.~I.}\ \bibnamefont {Mazin}}, \bibinfo {author}
  {\bibfnamefont {Y.}~\bibnamefont {Li}}, \bibinfo {author} {\bibfnamefont
  {H.~O.}\ \bibnamefont {Jeschke}}, \bibinfo {author} {\bibfnamefont
  {R.}~\bibnamefont {Valent\'{\i}}}, \ and\ \bibinfo {author} {\bibfnamefont
  {R.}~\bibnamefont {Coldea}},\ }\href {\doibase 10.1103/PhysRevB.92.235119}
  {\bibfield  {journal} {\bibinfo  {journal} {Phys. Rev. B}\ }\textbf {\bibinfo
  {volume} {92}},\ \bibinfo {pages} {235119} (\bibinfo {year}
  {2015})}\BibitemShut {NoStop}%
\bibitem [{\citenamefont {Kim}\ and\ \citenamefont {Kee}(2016)}]{HSKim2016PRB}%
  \BibitemOpen
  \bibfield  {author} {\bibinfo {author} {\bibfnamefont {H.-S.}\ \bibnamefont
  {Kim}}\ and\ \bibinfo {author} {\bibfnamefont {H.-Y.}\ \bibnamefont {Kee}},\
  }\href {\doibase 10.1103/PhysRevB.93.155143} {\bibfield  {journal} {\bibinfo
  {journal} {Phys. Rev. B}\ }\textbf {\bibinfo {volume} {93}},\ \bibinfo
  {pages} {155143} (\bibinfo {year} {2016})}\BibitemShut {NoStop}%
\bibitem [{\citenamefont {Cao}\ \emph {et~al.}(2016)\citenamefont {Cao},
  \citenamefont {Banerjee}, \citenamefont {Yan}, \citenamefont {Bridges},
  \citenamefont {Lumsden}, \citenamefont {Mandrus}, \citenamefont {Tennant},
  \citenamefont {Chakoumakos},\ and\ \citenamefont {Nagler}}]{Cao2016PRB}%
  \BibitemOpen
  \bibfield  {author} {\bibinfo {author} {\bibfnamefont {H.~B.}\ \bibnamefont
  {Cao}}, \bibinfo {author} {\bibfnamefont {A.}~\bibnamefont {Banerjee}},
  \bibinfo {author} {\bibfnamefont {J.-Q.}\ \bibnamefont {Yan}}, \bibinfo
  {author} {\bibfnamefont {C.~A.}\ \bibnamefont {Bridges}}, \bibinfo {author}
  {\bibfnamefont {M.~D.}\ \bibnamefont {Lumsden}}, \bibinfo {author}
  {\bibfnamefont {D.~G.}\ \bibnamefont {Mandrus}}, \bibinfo {author}
  {\bibfnamefont {D.~A.}\ \bibnamefont {Tennant}}, \bibinfo {author}
  {\bibfnamefont {B.~C.}\ \bibnamefont {Chakoumakos}}, \ and\ \bibinfo {author}
  {\bibfnamefont {S.~E.}\ \bibnamefont {Nagler}},\ }\href {\doibase
  10.1103/PhysRevB.93.134423} {\bibfield  {journal} {\bibinfo  {journal} {Phys.
  Rev. B}\ }\textbf {\bibinfo {volume} {93}},\ \bibinfo {pages} {134423}
  (\bibinfo {year} {2016})}\BibitemShut {NoStop}%
\bibitem [{\citenamefont {Janssen}\ \emph {et~al.}(2017)\citenamefont
  {Janssen}, \citenamefont {Andrade},\ and\ \citenamefont
  {Vojta}}]{Janssen2017PRB}%
  \BibitemOpen
  \bibfield  {author} {\bibinfo {author} {\bibfnamefont {L.}~\bibnamefont
  {Janssen}}, \bibinfo {author} {\bibfnamefont {E.~C.}\ \bibnamefont
  {Andrade}}, \ and\ \bibinfo {author} {\bibfnamefont {M.}~\bibnamefont
  {Vojta}},\ }\href {\doibase 10.1103/PhysRevB.96.064430} {\bibfield  {journal}
  {\bibinfo  {journal} {Phys. Rev. B}\ }\textbf {\bibinfo {volume} {96}},\
  \bibinfo {pages} {064430} (\bibinfo {year} {2017})}\BibitemShut {NoStop}%
\bibitem [{\citenamefont {Winter}\ \emph {et~al.}(2017)\citenamefont {Winter},
  \citenamefont {Tsirlin}, \citenamefont {Daghofer}, \citenamefont {van~den
  Brink}, \citenamefont {Singh}, \citenamefont {Gegenwart},\ and\ \citenamefont
  {Valent{\'{\i}}}}]{Winter2017IOP}%
  \BibitemOpen
  \bibfield  {author} {\bibinfo {author} {\bibfnamefont {S.~M.}\ \bibnamefont
  {Winter}}, \bibinfo {author} {\bibfnamefont {A.~A.}\ \bibnamefont {Tsirlin}},
  \bibinfo {author} {\bibfnamefont {M.}~\bibnamefont {Daghofer}}, \bibinfo
  {author} {\bibfnamefont {J.}~\bibnamefont {van~den Brink}}, \bibinfo {author}
  {\bibfnamefont {Y.}~\bibnamefont {Singh}}, \bibinfo {author} {\bibfnamefont
  {P.}~\bibnamefont {Gegenwart}}, \ and\ \bibinfo {author} {\bibfnamefont
  {R.}~\bibnamefont {Valent{\'{\i}}}},\ }\href {\doibase
  10.1088/1361-648x/aa8cf5} {\bibfield  {journal} {\bibinfo  {journal} {Journal
  of Physics: Condensed Matter}\ }\textbf {\bibinfo {volume} {29}},\ \bibinfo
  {pages} {493002} (\bibinfo {year} {2017})}\BibitemShut {NoStop}%
\bibitem [{\citenamefont {Baskaran}\ \emph {et~al.}(2008)\citenamefont
  {Baskaran}, \citenamefont {Sen},\ and\ \citenamefont
  {Shankar}}]{Baskaran2008PRB}%
  \BibitemOpen
  \bibfield  {author} {\bibinfo {author} {\bibfnamefont {G.}~\bibnamefont
  {Baskaran}}, \bibinfo {author} {\bibfnamefont {D.}~\bibnamefont {Sen}}, \
  and\ \bibinfo {author} {\bibfnamefont {R.}~\bibnamefont {Shankar}},\ }\href
  {\doibase 10.1103/PhysRevB.78.115116} {\bibfield  {journal} {\bibinfo
  {journal} {Phys. Rev. B}\ }\textbf {\bibinfo {volume} {78}},\ \bibinfo
  {pages} {115116} (\bibinfo {year} {2008})}\BibitemShut {NoStop}%
\bibitem [{\citenamefont {Baskaran}\ \emph {et~al.}(2007)\citenamefont
  {Baskaran}, \citenamefont {Mandal},\ and\ \citenamefont
  {Shankar}}]{Baskaran2007PRL}%
  \BibitemOpen
  \bibfield  {author} {\bibinfo {author} {\bibfnamefont {G.}~\bibnamefont
  {Baskaran}}, \bibinfo {author} {\bibfnamefont {S.}~\bibnamefont {Mandal}}, \
  and\ \bibinfo {author} {\bibfnamefont {R.}~\bibnamefont {Shankar}},\ }\href
  {\doibase 10.1103/PhysRevLett.98.247201} {\bibfield  {journal} {\bibinfo
  {journal} {Phys. Rev. Lett.}\ }\textbf {\bibinfo {volume} {98}},\ \bibinfo
  {pages} {247201} (\bibinfo {year} {2007})}\BibitemShut {NoStop}%
\bibitem [{\citenamefont {Koga}\ \emph {et~al.}(2018)\citenamefont {Koga},
  \citenamefont {Tomishige},\ and\ \citenamefont {Nasu}}]{Nasu2018JPSJ}%
  \BibitemOpen
  \bibfield  {author} {\bibinfo {author} {\bibfnamefont {A.}~\bibnamefont
  {Koga}}, \bibinfo {author} {\bibfnamefont {H.}~\bibnamefont {Tomishige}}, \
  and\ \bibinfo {author} {\bibfnamefont {J.}~\bibnamefont {Nasu}},\ }\href
  {\doibase 10.7566/JPSJ.87.063703} {\bibfield  {journal} {\bibinfo  {journal}
  {Journal of the Physical Society of Japan}\ }\textbf {\bibinfo {volume}
  {87}},\ \bibinfo {pages} {063703} (\bibinfo {year} {2018})}\BibitemShut
  {NoStop}%
\bibitem [{\citenamefont {Oitmaa}\ \emph {et~al.}(2018)\citenamefont {Oitmaa},
  \citenamefont {Koga},\ and\ \citenamefont {Singh}}]{Singh2018PRB}%
  \BibitemOpen
  \bibfield  {author} {\bibinfo {author} {\bibfnamefont {J.}~\bibnamefont
  {Oitmaa}}, \bibinfo {author} {\bibfnamefont {A.}~\bibnamefont {Koga}}, \ and\
  \bibinfo {author} {\bibfnamefont {R.~R.~P.}\ \bibnamefont {Singh}},\ }\href
  {\doibase 10.1103/PhysRevB.98.214404} {\bibfield  {journal} {\bibinfo
  {journal} {Phys. Rev. B}\ }\textbf {\bibinfo {volume} {98}},\ \bibinfo
  {pages} {214404} (\bibinfo {year} {2018})}\BibitemShut {NoStop}%
\bibitem [{\citenamefont {Samarakoon}\ \emph {et~al.}(2017)\citenamefont
  {Samarakoon}, \citenamefont {Banerjee}, \citenamefont {Zhang}, \citenamefont
  {Kamiya}, \citenamefont {Nagler}, \citenamefont {Tennant}, \citenamefont
  {Lee},\ and\ \citenamefont {Batista}}]{BatistaPRB2017}%
  \BibitemOpen
  \bibfield  {author} {\bibinfo {author} {\bibfnamefont {A.~M.}\ \bibnamefont
  {Samarakoon}}, \bibinfo {author} {\bibfnamefont {A.}~\bibnamefont
  {Banerjee}}, \bibinfo {author} {\bibfnamefont {S.-S.}\ \bibnamefont {Zhang}},
  \bibinfo {author} {\bibfnamefont {Y.}~\bibnamefont {Kamiya}}, \bibinfo
  {author} {\bibfnamefont {S.~E.}\ \bibnamefont {Nagler}}, \bibinfo {author}
  {\bibfnamefont {D.~A.}\ \bibnamefont {Tennant}}, \bibinfo {author}
  {\bibfnamefont {S.-H.}\ \bibnamefont {Lee}}, \ and\ \bibinfo {author}
  {\bibfnamefont {C.~D.}\ \bibnamefont {Batista}},\ }\href {\doibase
  10.1103/PhysRevB.96.134408} {\bibfield  {journal} {\bibinfo  {journal} {Phys.
  Rev. B}\ }\textbf {\bibinfo {volume} {96}},\ \bibinfo {pages} {134408}
  (\bibinfo {year} {2017})}\BibitemShut {NoStop}%
\bibitem [{\citenamefont {Kanamori}(1963)}]{Kanamori1963PoTP}%
  \BibitemOpen
  \bibfield  {author} {\bibinfo {author} {\bibfnamefont {J.}~\bibnamefont
  {Kanamori}},\ }\href {\doibase 10.1143/PTP.30.275} {\bibfield  {journal}
  {\bibinfo  {journal} {Progress of Theoretical Physics}\ }\textbf {\bibinfo
  {volume} {30}},\ \bibinfo {pages} {275} (\bibinfo {year} {1963})}\BibitemShut
  {NoStop}%
\bibitem [{\citenamefont {Fazekas}(1999)}]{fazekas1999}%
  \BibitemOpen
  \bibfield  {author} {\bibinfo {author} {\bibfnamefont {P.}~\bibnamefont
  {Fazekas}},\ }\enquote {\bibinfo {title} {Lecture notes on electron
  correlation and magnetism},}\ \ (\bibinfo  {publisher} {World Scientific},\
  \bibinfo {year} {1999})\BibitemShut {NoStop}%
\bibitem [{\citenamefont {Slater}\ and\ \citenamefont
  {Koster}(1954)}]{SlaterKoster}%
  \BibitemOpen
  \bibfield  {author} {\bibinfo {author} {\bibfnamefont {J.~C.}\ \bibnamefont
  {Slater}}\ and\ \bibinfo {author} {\bibfnamefont {G.~F.}\ \bibnamefont
  {Koster}},\ }\href {\doibase 10.1103/PhysRev.94.1498} {\bibfield  {journal}
  {\bibinfo  {journal} {Phys. Rev.}\ }\textbf {\bibinfo {volume} {94}},\
  \bibinfo {pages} {1498} (\bibinfo {year} {1954})}\BibitemShut {NoStop}%
\bibitem [{\citenamefont {Yamaji}\ \emph {et~al.}(2014)\citenamefont {Yamaji},
  \citenamefont {Nomura}, \citenamefont {Kurita}, \citenamefont {Arita},\ and\
  \citenamefont {Imada}}]{Imada2014PRL}%
  \BibitemOpen
  \bibfield  {author} {\bibinfo {author} {\bibfnamefont {Y.}~\bibnamefont
  {Yamaji}}, \bibinfo {author} {\bibfnamefont {Y.}~\bibnamefont {Nomura}},
  \bibinfo {author} {\bibfnamefont {M.}~\bibnamefont {Kurita}}, \bibinfo
  {author} {\bibfnamefont {R.}~\bibnamefont {Arita}}, \ and\ \bibinfo {author}
  {\bibfnamefont {M.}~\bibnamefont {Imada}},\ }\href {\doibase
  10.1103/PhysRevLett.113.107201} {\bibfield  {journal} {\bibinfo  {journal}
  {Phys. Rev. Lett.}\ }\textbf {\bibinfo {volume} {113}},\ \bibinfo {pages}
  {107201} (\bibinfo {year} {2014})}\BibitemShut {NoStop}%
\bibitem [{\citenamefont {Katukuri}\ \emph {et~al.}(2014)\citenamefont
  {Katukuri}, \citenamefont {Nishimoto}, \citenamefont {Yushankhai},
  \citenamefont {Stoyanova}, \citenamefont {Kandpal}, \citenamefont {Choi},
  \citenamefont {Coldea}, \citenamefont {Rousochatzakis}, \citenamefont
  {Hozoi},\ and\ \citenamefont {van~den Brink}}]{VdBrink2014IOP}%
  \BibitemOpen
  \bibfield  {author} {\bibinfo {author} {\bibfnamefont {V.~M.}\ \bibnamefont
  {Katukuri}}, \bibinfo {author} {\bibfnamefont {S.}~\bibnamefont {Nishimoto}},
  \bibinfo {author} {\bibfnamefont {V.}~\bibnamefont {Yushankhai}}, \bibinfo
  {author} {\bibfnamefont {A.}~\bibnamefont {Stoyanova}}, \bibinfo {author}
  {\bibfnamefont {H.}~\bibnamefont {Kandpal}}, \bibinfo {author} {\bibfnamefont
  {S.}~\bibnamefont {Choi}}, \bibinfo {author} {\bibfnamefont {R.}~\bibnamefont
  {Coldea}}, \bibinfo {author} {\bibfnamefont {I.}~\bibnamefont
  {Rousochatzakis}}, \bibinfo {author} {\bibfnamefont {L.}~\bibnamefont
  {Hozoi}}, \ and\ \bibinfo {author} {\bibfnamefont {J.}~\bibnamefont {van~den
  Brink}},\ }\href {\doibase 10.1088/1367-2630/16/1/013056} {\bibfield
  {journal} {\bibinfo  {journal} {New Journal of Physics}\ }\textbf {\bibinfo
  {volume} {16}},\ \bibinfo {pages} {013056} (\bibinfo {year}
  {2014})}\BibitemShut {NoStop}%
\bibitem [{\citenamefont {Busey}\ and\ \citenamefont
  {Giauque}(1952)}]{BuseyJoACS1952}%
  \BibitemOpen
  \bibfield  {author} {\bibinfo {author} {\bibfnamefont {R.~H.}\ \bibnamefont
  {Busey}}\ and\ \bibinfo {author} {\bibfnamefont {W.~F.}\ \bibnamefont
  {Giauque}},\ }\href {\doibase 10.1021/ja01137a062} {\bibfield  {journal}
  {\bibinfo  {journal} {Journal of the American Chemical Society}\ }\textbf
  {\bibinfo {volume} {74}},\ \bibinfo {pages} {4443} (\bibinfo {year}
  {1952})}\BibitemShut {NoStop}%
\bibitem [{\citenamefont {Billerey}\ \emph {et~al.}(1977)\citenamefont
  {Billerey}, \citenamefont {Terrier}, \citenamefont {Ciret},\ and\
  \citenamefont {Kleinclauss}}]{Billerey1977PLA}%
  \BibitemOpen
  \bibfield  {author} {\bibinfo {author} {\bibfnamefont {D.}~\bibnamefont
  {Billerey}}, \bibinfo {author} {\bibfnamefont {C.}~\bibnamefont {Terrier}},
  \bibinfo {author} {\bibfnamefont {N.}~\bibnamefont {Ciret}}, \ and\ \bibinfo
  {author} {\bibfnamefont {J.}~\bibnamefont {Kleinclauss}},\ }\href {\doibase
  https://doi.org/10.1016/0375-9601(77)90863-5} {\bibfield  {journal} {\bibinfo
   {journal} {Physics Letters A}\ }\textbf {\bibinfo {volume} {61}},\ \bibinfo
  {pages} {138 } (\bibinfo {year} {1977})}\BibitemShut {NoStop}%
\bibitem [{\citenamefont {Kuindersma}\ \emph {et~al.}(1981)\citenamefont
  {Kuindersma}, \citenamefont {Sanchez},\ and\ \citenamefont
  {Haas}}]{KuindersmaPhysica1981}%
  \BibitemOpen
  \bibfield  {author} {\bibinfo {author} {\bibfnamefont {S.}~\bibnamefont
  {Kuindersma}}, \bibinfo {author} {\bibfnamefont {J.}~\bibnamefont {Sanchez}},
  \ and\ \bibinfo {author} {\bibfnamefont {C.}~\bibnamefont {Haas}},\ }\href
  {https://doi.org/10.1016/0378-4363(81)90100-5} {\bibfield  {journal}
  {\bibinfo  {journal} {Physica B+C}\ }\textbf {\bibinfo {volume} {111}},\
  \bibinfo {pages} {231 } (\bibinfo {year} {1981})}\BibitemShut {NoStop}%
\bibitem [{\citenamefont {McGuire}\ \emph {et~al.}(2015)\citenamefont
  {McGuire}, \citenamefont {Dixit}, \citenamefont {Cooper},\ and\ \citenamefont
  {Sales}}]{McGuire2015CoM}%
  \BibitemOpen
  \bibfield  {author} {\bibinfo {author} {\bibfnamefont {M.~A.}\ \bibnamefont
  {McGuire}}, \bibinfo {author} {\bibfnamefont {H.}~\bibnamefont {Dixit}},
  \bibinfo {author} {\bibfnamefont {V.~R.}\ \bibnamefont {Cooper}}, \ and\
  \bibinfo {author} {\bibfnamefont {B.~C.}\ \bibnamefont {Sales}},\ }\href
  {\doibase 10.1021/cm504242t} {\bibfield  {journal} {\bibinfo  {journal}
  {Chemistry of Materials}\ }\textbf {\bibinfo {volume} {27}},\ \bibinfo
  {pages} {612} (\bibinfo {year} {2015})}\BibitemShut {NoStop}%
\bibitem [{\citenamefont {Zvereva}\ \emph {et~al.}(2015)\citenamefont
  {Zvereva}, \citenamefont {Stratan}, \citenamefont {Ovchenkov}, \citenamefont
  {Nalbandyan}, \citenamefont {Lin}, \citenamefont {Vavilova}, \citenamefont
  {Iakovleva}, \citenamefont {Abdel-Hafiez}, \citenamefont {Silhanek},
  \citenamefont {Chen}, \citenamefont {Stroppa}, \citenamefont {Picozzi},
  \citenamefont {Jeschke}, \citenamefont {Valent\'{\i}},\ and\ \citenamefont
  {Vasiliev}}]{ZverevaPRB2015}%
  \BibitemOpen
  \bibfield  {author} {\bibinfo {author} {\bibfnamefont {E.~A.}\ \bibnamefont
  {Zvereva}}, \bibinfo {author} {\bibfnamefont {M.~I.}\ \bibnamefont
  {Stratan}}, \bibinfo {author} {\bibfnamefont {Y.~A.}\ \bibnamefont
  {Ovchenkov}}, \bibinfo {author} {\bibfnamefont {V.~B.}\ \bibnamefont
  {Nalbandyan}}, \bibinfo {author} {\bibfnamefont {J.-Y.}\ \bibnamefont {Lin}},
  \bibinfo {author} {\bibfnamefont {E.~L.}\ \bibnamefont {Vavilova}}, \bibinfo
  {author} {\bibfnamefont {M.~F.}\ \bibnamefont {Iakovleva}}, \bibinfo {author}
  {\bibfnamefont {M.}~\bibnamefont {Abdel-Hafiez}}, \bibinfo {author}
  {\bibfnamefont {A.~V.}\ \bibnamefont {Silhanek}}, \bibinfo {author}
  {\bibfnamefont {X.-J.}\ \bibnamefont {Chen}}, \bibinfo {author}
  {\bibfnamefont {A.}~\bibnamefont {Stroppa}}, \bibinfo {author} {\bibfnamefont
  {S.}~\bibnamefont {Picozzi}}, \bibinfo {author} {\bibfnamefont {H.~O.}\
  \bibnamefont {Jeschke}}, \bibinfo {author} {\bibfnamefont {R.}~\bibnamefont
  {Valent\'{\i}}}, \ and\ \bibinfo {author} {\bibfnamefont {A.~N.}\
  \bibnamefont {Vasiliev}},\ }\href {\doibase 10.1103/PhysRevB.92.144401}
  {\bibfield  {journal} {\bibinfo  {journal} {Phys. Rev. B}\ }\textbf {\bibinfo
  {volume} {92}},\ \bibinfo {pages} {144401} (\bibinfo {year}
  {2015})}\BibitemShut {NoStop}%
\bibitem [{\citenamefont {Kurbakov}\ \emph {et~al.}(2017)\citenamefont
  {Kurbakov}, \citenamefont {Korshunov}, \citenamefont {Podchezertsev},
  \citenamefont {Malyshev}, \citenamefont {Evstigneeva}, \citenamefont {Damay},
  \citenamefont {Park}, \citenamefont {Koo}, \citenamefont {Klingeler},
  \citenamefont {Zvereva},\ and\ \citenamefont
  {Nalbandyan}}]{NalbandyanPRB2017}%
  \BibitemOpen
  \bibfield  {author} {\bibinfo {author} {\bibfnamefont {A.~I.}\ \bibnamefont
  {Kurbakov}}, \bibinfo {author} {\bibfnamefont {A.~N.}\ \bibnamefont
  {Korshunov}}, \bibinfo {author} {\bibfnamefont {S.~Y.}\ \bibnamefont
  {Podchezertsev}}, \bibinfo {author} {\bibfnamefont {A.~L.}\ \bibnamefont
  {Malyshev}}, \bibinfo {author} {\bibfnamefont {M.~A.}\ \bibnamefont
  {Evstigneeva}}, \bibinfo {author} {\bibfnamefont {F.}~\bibnamefont {Damay}},
  \bibinfo {author} {\bibfnamefont {J.}~\bibnamefont {Park}}, \bibinfo {author}
  {\bibfnamefont {C.}~\bibnamefont {Koo}}, \bibinfo {author} {\bibfnamefont
  {R.}~\bibnamefont {Klingeler}}, \bibinfo {author} {\bibfnamefont {E.~A.}\
  \bibnamefont {Zvereva}}, \ and\ \bibinfo {author} {\bibfnamefont {V.~B.}\
  \bibnamefont {Nalbandyan}},\ }\href {\doibase 10.1103/PhysRevB.96.024417}
  {\bibfield  {journal} {\bibinfo  {journal} {Phys. Rev. B}\ }\textbf {\bibinfo
  {volume} {96}},\ \bibinfo {pages} {024417} (\bibinfo {year}
  {2017})}\BibitemShut {NoStop}%
\bibitem [{\citenamefont {Dillon}\ and\ \citenamefont
  {Olson}(1965)}]{Dillon1965JoAP}%
  \BibitemOpen
  \bibfield  {author} {\bibinfo {author} {\bibfnamefont {J.~F.}\ \bibnamefont
  {Dillon}}\ and\ \bibinfo {author} {\bibfnamefont {C.~E.}\ \bibnamefont
  {Olson}},\ }\href {\doibase 10.1063/1.1714194} {\bibfield  {journal}
  {\bibinfo  {journal} {Journal of Applied Physics}\ }\textbf {\bibinfo
  {volume} {36}},\ \bibinfo {pages} {1259} (\bibinfo {year}
  {1965})}\BibitemShut {NoStop}%
\bibitem [{\citenamefont {Huang}\ \emph {et~al.}(2017)\citenamefont {Huang},
  \citenamefont {Clark}, \citenamefont {Navarro-Moratalla}, \citenamefont
  {Klein}, \citenamefont {Cheng}, \citenamefont {Seyler}, \citenamefont
  {Zhong}, \citenamefont {Schmidgall}, \citenamefont {McGuire}, \citenamefont
  {Cobden}, \citenamefont {Yao}, \citenamefont {Xiao}, \citenamefont
  {Jarillo-Herrero},\ and\ \citenamefont {Xu}}]{Huang2017Nature}%
  \BibitemOpen
  \bibfield  {author} {\bibinfo {author} {\bibfnamefont {B.}~\bibnamefont
  {Huang}}, \bibinfo {author} {\bibfnamefont {G.}~\bibnamefont {Clark}},
  \bibinfo {author} {\bibfnamefont {E.}~\bibnamefont {Navarro-Moratalla}},
  \bibinfo {author} {\bibfnamefont {D.~R.}\ \bibnamefont {Klein}}, \bibinfo
  {author} {\bibfnamefont {R.}~\bibnamefont {Cheng}}, \bibinfo {author}
  {\bibfnamefont {K.~L.}\ \bibnamefont {Seyler}}, \bibinfo {author}
  {\bibfnamefont {D.}~\bibnamefont {Zhong}}, \bibinfo {author} {\bibfnamefont
  {E.}~\bibnamefont {Schmidgall}}, \bibinfo {author} {\bibfnamefont {M.~A.}\
  \bibnamefont {McGuire}}, \bibinfo {author} {\bibfnamefont {D.~H.}\
  \bibnamefont {Cobden}}, \bibinfo {author} {\bibfnamefont {W.}~\bibnamefont
  {Yao}}, \bibinfo {author} {\bibfnamefont {D.}~\bibnamefont {Xiao}}, \bibinfo
  {author} {\bibfnamefont {P.}~\bibnamefont {Jarillo-Herrero}}, \ and\ \bibinfo
  {author} {\bibfnamefont {X.}~\bibnamefont {Xu}},\ }\href {\doibase
  https://doi.org/10.1038/nature22391} {\bibfield  {journal} {\bibinfo
  {journal} {Nature}\ }\textbf {\bibinfo {volume} {546}},\ \bibinfo {pages}
  {270} (\bibinfo {year} {2017})}\BibitemShut {NoStop}%
\bibitem [{\citenamefont {McGuire}(2017)}]{McGuire2017Crystals}%
  \BibitemOpen
  \bibfield  {author} {\bibinfo {author} {\bibfnamefont {M.~A.}\ \bibnamefont
  {McGuire}},\ }\href {\doibase 10.3390/cryst7050121} {\bibfield  {journal}
  {\bibinfo  {journal} {Crystals}\ }\textbf {\bibinfo {volume} {7}} (\bibinfo
  {year} {2017}),\ 10.3390/cryst7050121}\BibitemShut {NoStop}%
\bibitem [{\citenamefont {Yadav}\ \emph {et~al.}(2016)\citenamefont {Yadav},
  \citenamefont {Bogdanov}, \citenamefont {Katukuri}, \citenamefont
  {Nishimoto}, \citenamefont {Van Den~Brink},\ and\ \citenamefont
  {Hozoi}}]{Yadav2016SCIR}%
  \BibitemOpen
  \bibfield  {author} {\bibinfo {author} {\bibfnamefont {R.}~\bibnamefont
  {Yadav}}, \bibinfo {author} {\bibfnamefont {N.~A.}\ \bibnamefont {Bogdanov}},
  \bibinfo {author} {\bibfnamefont {V.~M.}\ \bibnamefont {Katukuri}}, \bibinfo
  {author} {\bibfnamefont {S.}~\bibnamefont {Nishimoto}}, \bibinfo {author}
  {\bibfnamefont {J.}~\bibnamefont {Van Den~Brink}}, \ and\ \bibinfo {author}
  {\bibfnamefont {L.}~\bibnamefont {Hozoi}},\ }\href
  {https://www.ncbi.nlm.nih.gov/pubmed/27901091} {\bibfield  {journal}
  {\bibinfo  {journal} {Scientific Reports}\ }\textbf {\bibinfo {volume} {6}},\
  \bibinfo {pages} {37925} (\bibinfo {year} {2016})}\BibitemShut {NoStop}%
\bibitem [{\citenamefont {Baek}\ \emph {et~al.}(2017)\citenamefont {Baek},
  \citenamefont {Do}, \citenamefont {Choi}, \citenamefont {Kwon}, \citenamefont
  {Wolter}, \citenamefont {Nishimoto}, \citenamefont {van~den Brink},\ and\
  \citenamefont {B\"uchner}}]{Baek2017PRL}%
  \BibitemOpen
  \bibfield  {author} {\bibinfo {author} {\bibfnamefont {S.-H.}\ \bibnamefont
  {Baek}}, \bibinfo {author} {\bibfnamefont {S.-H.}\ \bibnamefont {Do}},
  \bibinfo {author} {\bibfnamefont {K.-Y.}\ \bibnamefont {Choi}}, \bibinfo
  {author} {\bibfnamefont {Y.~S.}\ \bibnamefont {Kwon}}, \bibinfo {author}
  {\bibfnamefont {A.~U.~B.}\ \bibnamefont {Wolter}}, \bibinfo {author}
  {\bibfnamefont {S.}~\bibnamefont {Nishimoto}}, \bibinfo {author}
  {\bibfnamefont {J.}~\bibnamefont {van~den Brink}}, \ and\ \bibinfo {author}
  {\bibfnamefont {B.}~\bibnamefont {B\"uchner}},\ }\href {\doibase
  10.1103/PhysRevLett.119.037201} {\bibfield  {journal} {\bibinfo  {journal}
  {Phys. Rev. Lett.}\ }\textbf {\bibinfo {volume} {119}},\ \bibinfo {pages}
  {037201} (\bibinfo {year} {2017})}\BibitemShut {NoStop}%
\bibitem [{\citenamefont {Wolter}\ \emph {et~al.}(2017)\citenamefont {Wolter},
  \citenamefont {Corredor}, \citenamefont {Janssen}, \citenamefont {Nenkov},
  \citenamefont {Sch\"onecker}, \citenamefont {Do}, \citenamefont {Choi},
  \citenamefont {Albrecht}, \citenamefont {Hunger}, \citenamefont {Doert},
  \citenamefont {Vojta},\ and\ \citenamefont {B\"uchner}}]{Wolter2017PRB}%
  \BibitemOpen
  \bibfield  {author} {\bibinfo {author} {\bibfnamefont {A.~U.~B.}\
  \bibnamefont {Wolter}}, \bibinfo {author} {\bibfnamefont {L.~T.}\
  \bibnamefont {Corredor}}, \bibinfo {author} {\bibfnamefont {L.}~\bibnamefont
  {Janssen}}, \bibinfo {author} {\bibfnamefont {K.}~\bibnamefont {Nenkov}},
  \bibinfo {author} {\bibfnamefont {S.}~\bibnamefont {Sch\"onecker}}, \bibinfo
  {author} {\bibfnamefont {S.-H.}\ \bibnamefont {Do}}, \bibinfo {author}
  {\bibfnamefont {K.-Y.}\ \bibnamefont {Choi}}, \bibinfo {author}
  {\bibfnamefont {R.}~\bibnamefont {Albrecht}}, \bibinfo {author}
  {\bibfnamefont {J.}~\bibnamefont {Hunger}}, \bibinfo {author} {\bibfnamefont
  {T.}~\bibnamefont {Doert}}, \bibinfo {author} {\bibfnamefont
  {M.}~\bibnamefont {Vojta}}, \ and\ \bibinfo {author} {\bibfnamefont
  {B.}~\bibnamefont {B\"uchner}},\ }\href {\doibase 10.1103/PhysRevB.96.041405}
  {\bibfield  {journal} {\bibinfo  {journal} {Phys. Rev. B}\ }\textbf {\bibinfo
  {volume} {96}},\ \bibinfo {pages} {041405} (\bibinfo {year}
  {2017})}\BibitemShut {NoStop}%
\bibitem [{\citenamefont {Zheng}\ \emph {et~al.}(2017)\citenamefont {Zheng},
  \citenamefont {Ran}, \citenamefont {Li}, \citenamefont {Wang}, \citenamefont
  {Wang}, \citenamefont {Liu}, \citenamefont {Liu}, \citenamefont {Normand},
  \citenamefont {Wen},\ and\ \citenamefont {Yu}}]{Zheng2017PRL}%
  \BibitemOpen
  \bibfield  {author} {\bibinfo {author} {\bibfnamefont {J.}~\bibnamefont
  {Zheng}}, \bibinfo {author} {\bibfnamefont {K.}~\bibnamefont {Ran}}, \bibinfo
  {author} {\bibfnamefont {T.}~\bibnamefont {Li}}, \bibinfo {author}
  {\bibfnamefont {J.}~\bibnamefont {Wang}}, \bibinfo {author} {\bibfnamefont
  {P.}~\bibnamefont {Wang}}, \bibinfo {author} {\bibfnamefont {B.}~\bibnamefont
  {Liu}}, \bibinfo {author} {\bibfnamefont {Z.-X.}\ \bibnamefont {Liu}},
  \bibinfo {author} {\bibfnamefont {B.}~\bibnamefont {Normand}}, \bibinfo
  {author} {\bibfnamefont {J.}~\bibnamefont {Wen}}, \ and\ \bibinfo {author}
  {\bibfnamefont {W.}~\bibnamefont {Yu}},\ }\href {\doibase
  10.1103/PhysRevLett.119.227208} {\bibfield  {journal} {\bibinfo  {journal}
  {Phys. Rev. Lett.}\ }\textbf {\bibinfo {volume} {119}},\ \bibinfo {pages}
  {227208} (\bibinfo {year} {2017})}\BibitemShut {NoStop}%
\bibitem [{\citenamefont {Jansa}\ \emph {et~al.}(2018)\citenamefont {Jansa},
  \citenamefont {Zorko}, \citenamefont {Gomilsek}, \citenamefont {Pregelj},
  \citenamefont {Kr{\"a}mer}, \citenamefont {Biner}, \citenamefont {Biffin},
  \citenamefont {R{\"u}egg},\ and\ \citenamefont {Klanjsek}}]{Jansa2018NAP}%
  \BibitemOpen
  \bibfield  {author} {\bibinfo {author} {\bibfnamefont {N.}~\bibnamefont
  {Jansa}}, \bibinfo {author} {\bibfnamefont {A.}~\bibnamefont {Zorko}},
  \bibinfo {author} {\bibfnamefont {M.}~\bibnamefont {Gomilsek}}, \bibinfo
  {author} {\bibfnamefont {M.}~\bibnamefont {Pregelj}}, \bibinfo {author}
  {\bibfnamefont {K.~W.}\ \bibnamefont {Kr{\"a}mer}}, \bibinfo {author}
  {\bibfnamefont {D.}~\bibnamefont {Biner}}, \bibinfo {author} {\bibfnamefont
  {A.}~\bibnamefont {Biffin}}, \bibinfo {author} {\bibfnamefont
  {C.}~\bibnamefont {R{\"u}egg}}, \ and\ \bibinfo {author} {\bibfnamefont
  {M.}~\bibnamefont {Klanjsek}},\ }\href {\doibase 10.1038/s41567-018-0129-5}
  {\bibfield  {journal} {\bibinfo  {journal} {Nature Physics}\ }\textbf
  {\bibinfo {volume} {14}},\ \bibinfo {pages} {786} (\bibinfo {year}
  {2018})}\BibitemShut {NoStop}%
\bibitem [{\citenamefont {Zhu}\ \emph {et~al.}(2018)\citenamefont {Zhu},
  \citenamefont {Kimchi}, \citenamefont {Sheng},\ and\ \citenamefont
  {Fu}}]{Fu2018PRB}%
  \BibitemOpen
  \bibfield  {author} {\bibinfo {author} {\bibfnamefont {Z.}~\bibnamefont
  {Zhu}}, \bibinfo {author} {\bibfnamefont {I.}~\bibnamefont {Kimchi}},
  \bibinfo {author} {\bibfnamefont {D.~N.}\ \bibnamefont {Sheng}}, \ and\
  \bibinfo {author} {\bibfnamefont {L.}~\bibnamefont {Fu}},\ }\href {\doibase
  10.1103/PhysRevB.97.241110} {\bibfield  {journal} {\bibinfo  {journal} {Phys.
  Rev. B}\ }\textbf {\bibinfo {volume} {97}},\ \bibinfo {pages} {241110}
  (\bibinfo {year} {2018})}\BibitemShut {NoStop}%
\bibitem [{\citenamefont {Gohlke}\ \emph {et~al.}(2018)\citenamefont {Gohlke},
  \citenamefont {Moessner},\ and\ \citenamefont {Pollmann}}]{GohlkePRB2018}%
  \BibitemOpen
  \bibfield  {author} {\bibinfo {author} {\bibfnamefont {M.}~\bibnamefont
  {Gohlke}}, \bibinfo {author} {\bibfnamefont {R.}~\bibnamefont {Moessner}}, \
  and\ \bibinfo {author} {\bibfnamefont {F.}~\bibnamefont {Pollmann}},\ }\href
  {\doibase 10.1103/PhysRevB.98.014418} {\bibfield  {journal} {\bibinfo
  {journal} {Phys. Rev. B}\ }\textbf {\bibinfo {volume} {98}},\ \bibinfo
  {pages} {014418} (\bibinfo {year} {2018})}\BibitemShut {NoStop}%
\bibitem [{\citenamefont {Nasu}\ \emph {et~al.}(2018)\citenamefont {Nasu},
  \citenamefont {Kato}, \citenamefont {Kamiya},\ and\ \citenamefont
  {Motome}}]{NasuPRB2018}%
  \BibitemOpen
  \bibfield  {author} {\bibinfo {author} {\bibfnamefont {J.}~\bibnamefont
  {Nasu}}, \bibinfo {author} {\bibfnamefont {Y.}~\bibnamefont {Kato}}, \bibinfo
  {author} {\bibfnamefont {Y.}~\bibnamefont {Kamiya}}, \ and\ \bibinfo {author}
  {\bibfnamefont {Y.}~\bibnamefont {Motome}},\ }\href {\doibase
  10.1103/PhysRevB.98.060416} {\bibfield  {journal} {\bibinfo  {journal} {Phys.
  Rev. B}\ }\textbf {\bibinfo {volume} {98}},\ \bibinfo {pages} {060416}
  (\bibinfo {year} {2018})}\BibitemShut {NoStop}%
\bibitem [{\citenamefont {Ronquillo}\ \emph {et~al.}()\citenamefont
  {Ronquillo}, \citenamefont {Vengal},\ and\ \citenamefont
  {Trivedi}}]{Ronquillo2018ARXIV}%
  \BibitemOpen
  \bibfield  {author} {\bibinfo {author} {\bibfnamefont {D.~C.}\ \bibnamefont
  {Ronquillo}}, \bibinfo {author} {\bibfnamefont {A.}~\bibnamefont {Vengal}}, \
  and\ \bibinfo {author} {\bibfnamefont {N.}~\bibnamefont {Trivedi}},\
  }\href@noop {} {}\Eprint {http://arxiv.org/abs/arXiv:1805.03722}
  {arXiv:1805.03722} \BibitemShut {NoStop}%
\bibitem [{\citenamefont {Hickey}\ and\ \citenamefont
  {Trebst}(2019)}]{Ciaran2018NC}%
  \BibitemOpen
  \bibfield  {author} {\bibinfo {author} {\bibfnamefont {C.}~\bibnamefont
  {Hickey}}\ and\ \bibinfo {author} {\bibfnamefont {S.}~\bibnamefont
  {Trebst}},\ }\href {https://doi.org/10.1038/s41467-019-08459-9} {\bibfield
  {journal} {\bibinfo  {journal} {Nature Communications}\ }\textbf {\bibinfo
  {volume} {10}},\ \bibinfo {pages} {530} (\bibinfo {year} {2019})}\BibitemShut
  {NoStop}%
\bibitem [{\citenamefont {Liang}\ \emph {et~al.}(2018)\citenamefont {Liang},
  \citenamefont {Jiang}, \citenamefont {Chen}, \citenamefont {Li},\ and\
  \citenamefont {Wang}}]{LiangPRB2018}%
  \BibitemOpen
  \bibfield  {author} {\bibinfo {author} {\bibfnamefont {S.}~\bibnamefont
  {Liang}}, \bibinfo {author} {\bibfnamefont {M.-H.}\ \bibnamefont {Jiang}},
  \bibinfo {author} {\bibfnamefont {W.}~\bibnamefont {Chen}}, \bibinfo {author}
  {\bibfnamefont {J.-X.}\ \bibnamefont {Li}}, \ and\ \bibinfo {author}
  {\bibfnamefont {Q.-H.}\ \bibnamefont {Wang}},\ }\href {\doibase
  10.1103/PhysRevB.98.054433} {\bibfield  {journal} {\bibinfo  {journal} {Phys.
  Rev. B}\ }\textbf {\bibinfo {volume} {98}},\ \bibinfo {pages} {054433}
  (\bibinfo {year} {2018})}\BibitemShut {NoStop}%
\bibitem [{\citenamefont {Lampen-Kelley}\ \emph {et~al.}()\citenamefont
  {Lampen-Kelley}, \citenamefont {Janssen}, \citenamefont {Andrade},
  \citenamefont {Rachel}, \citenamefont {Yan}, \citenamefont {Balz},
  \citenamefont {Mandrus}, \citenamefont {Nagler},\ and\ \citenamefont
  {Vojta}}]{Lampenkelley2018ARXIV}%
  \BibitemOpen
  \bibfield  {author} {\bibinfo {author} {\bibfnamefont {P.}~\bibnamefont
  {Lampen-Kelley}}, \bibinfo {author} {\bibfnamefont {L.}~\bibnamefont
  {Janssen}}, \bibinfo {author} {\bibfnamefont {E.~C.}\ \bibnamefont
  {Andrade}}, \bibinfo {author} {\bibfnamefont {S.}~\bibnamefont {Rachel}},
  \bibinfo {author} {\bibfnamefont {J.~Q.}\ \bibnamefont {Yan}}, \bibinfo
  {author} {\bibfnamefont {C.}~\bibnamefont {Balz}}, \bibinfo {author}
  {\bibfnamefont {D.~G.}\ \bibnamefont {Mandrus}}, \bibinfo {author}
  {\bibfnamefont {S.~E.}\ \bibnamefont {Nagler}}, \ and\ \bibinfo {author}
  {\bibfnamefont {M.}~\bibnamefont {Vojta}},\ }\href@noop {} {\ }\Eprint
  {http://arxiv.org/abs/arXiv:1807.06192} {arXiv:1807.06192} \BibitemShut
  {NoStop}%
\bibitem [{\citenamefont {Jiang}\ \emph {et~al.}()\citenamefont {Jiang},
  \citenamefont {Wang}, \citenamefont {Huang},\ and\ \citenamefont
  {Lu}}]{Lu2018ARXIV}%
  \BibitemOpen
  \bibfield  {author} {\bibinfo {author} {\bibfnamefont {H.-C.}\ \bibnamefont
  {Jiang}}, \bibinfo {author} {\bibfnamefont {C.-Y.}\ \bibnamefont {Wang}},
  \bibinfo {author} {\bibfnamefont {B.}~\bibnamefont {Huang}}, \ and\ \bibinfo
  {author} {\bibfnamefont {Y.-M.}\ \bibnamefont {Lu}},\ }\href@noop {} {\
  }\Eprint {http://arxiv.org/abs/arXiv:1809.08247} {arXiv:1809.08247}
  \BibitemShut {NoStop}%
\bibitem [{\citenamefont {Zou}\ and\ \citenamefont {He}()}]{Zou2018ARXIV}%
  \BibitemOpen
  \bibfield  {author} {\bibinfo {author} {\bibfnamefont {L.}~\bibnamefont
  {Zou}}\ and\ \bibinfo {author} {\bibfnamefont {Y.-C.}\ \bibnamefont {He}},\
  }\href@noop {} {\ }\Eprint {http://arxiv.org/abs/arXiv:1809.09091}
  {arXiv:1809.09091} \BibitemShut {NoStop}%
\bibitem [{\citenamefont {Patel}\ and\ \citenamefont
  {Trivedi}()}]{Patel2018ARXIV}%
  \BibitemOpen
  \bibfield  {author} {\bibinfo {author} {\bibfnamefont {N.~D.}\ \bibnamefont
  {Patel}}\ and\ \bibinfo {author} {\bibfnamefont {N.}~\bibnamefont
  {Trivedi}},\ }\href@noop {} {\ }\Eprint
  {http://arxiv.org/abs/arXiv:1812.06105} {arXiv:1812.06105} \BibitemShut
  {NoStop}%
\bibitem [{\citenamefont {Gordon}\ \emph {et~al.}()\citenamefont {Gordon},
  \citenamefont {Catuneanu}, \citenamefont {S\o{}rensen},\ and\ \citenamefont
  {Kee}}]{Jacob2019ARXIV}%
  \BibitemOpen
  \bibfield  {author} {\bibinfo {author} {\bibfnamefont {J.~S.}\ \bibnamefont
  {Gordon}}, \bibinfo {author} {\bibfnamefont {A.}~\bibnamefont {Catuneanu}},
  \bibinfo {author} {\bibfnamefont {E.~S.}\ \bibnamefont {S\o{}rensen}}, \ and\
  \bibinfo {author} {\bibfnamefont {H.-Y.}\ \bibnamefont {Kee}},\ }\href@noop
  {} {\ }\Eprint {http://arxiv.org/abs/arXiv:1901.09943} {arXiv:1901.09943}
  \BibitemShut {NoStop}%
\bibitem [{\citenamefont {Lee}\ \emph {et~al.}()\citenamefont {Lee},
  \citenamefont {Utermohlen}, \citenamefont {Hwang}, \citenamefont {Weber},
  \citenamefont {Zhang}, \citenamefont {van Tol}, \citenamefont {Goldberger},
  \citenamefont {Trivedi},\ and\ \citenamefont {Hammel}}]{CrI3arxiv2019}%
  \BibitemOpen
  \bibfield  {author} {\bibinfo {author} {\bibfnamefont {I.}~\bibnamefont
  {Lee}}, \bibinfo {author} {\bibfnamefont {F.~G.}\ \bibnamefont {Utermohlen}},
  \bibinfo {author} {\bibfnamefont {K.}~\bibnamefont {Hwang}}, \bibinfo
  {author} {\bibfnamefont {D.}~\bibnamefont {Weber}}, \bibinfo {author}
  {\bibfnamefont {C.}~\bibnamefont {Zhang}}, \bibinfo {author} {\bibfnamefont
  {J.}~\bibnamefont {van Tol}}, \bibinfo {author} {\bibfnamefont {J.~E.}\
  \bibnamefont {Goldberger}}, \bibinfo {author} {\bibfnamefont
  {N.}~\bibnamefont {Trivedi}}, \ and\ \bibinfo {author} {\bibfnamefont
  {P.~C.}\ \bibnamefont {Hammel}},\ }\href@noop {} {\ }\Eprint
  {http://arxiv.org/abs/arXiv:1902.00077} {arXiv:1902.00077} \BibitemShut
  {NoStop}%
\end{thebibliography}%

\end{document}